# Evaluating the Plausible Range of N₂O Biosignatures on Exo-Earths: An Integrated Biogeochemical, Photochemical, and Spectral Modeling Approach

Edward W. Schwieterman[1,2,3,4], Stephanie L. Olson[2,5], Daria Pidhorodetska[1,2,3], Christopher T. Reinhard[2,3,6], Ainsley Ganti[7], Thomas J. Fauchez[3,8,9,10], Sandra T. Bastelberger[8,10,11,12], Jaime S. Crouse[8,10,11,13], Andy Ridgwell[1,2], and Timothy W. Lyons[1,2,3]

[1] Department of Earth and Planetary Sciences, University of California, Riverside, CA, USA; eschwiet@ucr.edu
[2] NASA Alternative Earths Team, Riverside, CA, USA
[3] NASA NExSS Virtual Planetary Laboratory Team, Seattle, WA, USA
[4] Blue Marble Space Institute of Science, Seattle, WA, USA
[5] Department of Earth, Atmospheric, and Planetary Science, Purdue University, West Lafayette, IN, USA
[6] School of Earth and Atmospheric Sciences, Georgia Institute of Technology, Atlanta, GA, USA
[7] The Potomac School, 1301 Potomac School Rd, McLean, VA 22101, USA
[8] NASA Goddard Space Flight Center, 8800 Greenbelt Road, Greenbelt, MD 20771, USA
[9] American University, Washington, DC, USA
[10] Sellers Exoplanet Environment Collaboration (SEEC), NASA GSFC, USA
[11] Department of Astronomy, University of Maryland, College Park, MD 20742, USA
[12] Center for Research and Exploration in Space Science and Technology, NASA/GSFC, Greenbelt, MD 20771, USA
[13] Southeastern Universities Research Association (SURA), 1201 New York Avenue NW, Washington, DC 20005, USA
*Received 2022 May 13; revised 2022 August 17; accepted 2022 August 24; published 2022 October 4*

## Abstract

Nitrous oxide (N₂O)—a product of microbial nitrogen metabolism—is a compelling exoplanet biosignature gas with distinctive spectral features in the near- and mid-infrared, and only minor abiotic sources on Earth. Previous investigations of N₂O as a biosignature have examined scenarios using Earthlike N₂O mixing ratios or surface fluxes, or those inferred from Earth's geologic record. However, biological fluxes of N₂O could be substantially higher, due to a lack of metal catalysts or if the last step of the denitrification metabolism that yields N₂ from N₂O had never evolved. Here, we use a global biogeochemical model coupled with photochemical and spectral models to systematically quantify the limits of plausible N₂O abundances and spectral detectability for Earth analogs orbiting main-sequence (FGKM) stars. We examine N₂O buildup over a range of oxygen conditions (1%–100% present atmospheric level) and N₂O fluxes (0.01–100 teramole per year; Tmol = $10^{12}$ mole) that are compatible with Earth's history. We find that N₂O fluxes of 10 [100] Tmol yr$^{-1}$ would lead to maximum N₂O abundances of ~5 [50] ppm for Earth–Sun analogs, 90 [1600] ppm for Earths around late K dwarfs, and 30 [300] ppm for an Earthlike TRAPPIST-1e. We simulate emission and transmission spectra for intermediate and maximum N₂O concentrations that are relevant to current and future space-based telescopes. We calculate the detectability of N₂O spectral features for high-flux scenarios for TRAPPIST-1e with JWST. We review potential false positives, including chemodenitrification and abiotic production via stellar activity, and identify key spectral and contextual discriminants to confirm or refute the biogenicity of the observed N₂O.

*Unified Astronomy Thesaurus concepts:* Astrobiology (74); Exoplanet atmospheres (487); Exoplanets (498); Habitable planets (695); Nitrous oxide (1114); Biosignatures (2018)

## 1. Introduction

To date, over 5000 exoplanetary systems have been discovered (Christiansen 2022),[14] including several planets that are rocky in composition and located within the circumstellar habitable zone of their host star (Kane et al. 2016; Kaltenegger et al. 2019). The James Webb Space Telescope (JWST) will allow us to probe the atmospheres of a small number of these temperate terrestrial exoplanets, such as the TRAPPIST-1 planets (Gillon et al. 2017; Luger et al. 2017; Morley et al. 2017; Lincowski et al. 2018; Fauchez et al. 2019; Lustig-Yaeger et al. 2019; Ducrot et al. 2020), while upcoming ground-based extremely large telescopes will facilitate the examination of nearby potentially habitable worlds, such as Proxima Centauri b (Anglada-Escudé et al. 2016; Ribas et al. 2016; Snellen et al. 2017; Meadows et al. 2018a). Ambitious future mission concepts, such as the IR/optical/UV observatory recommended by the 2020 Astronomy and Astrophysics Decadal Survey (National Academies of Sciences, Engineering, and Medicine 2021), or ESA's mid-IR (MIR) Large Interferometer for Exoplanets (LIFE) concept (Defrère et al. 2018; Quanz et al. 2018, 2022), would allow for the unprecedented atmospheric characterization of a larger number of temperate rocky planets orbiting stars in the solar neighborhood (most of which are yet to be discovered), though the observability of specific spectral features will be limited by the wavelength regime and observing mode.

One of the most compelling drivers of exoplanet science is the search for inhabited planets like Earth, which may be identified through remote spectroscopic biosignatures (Des Marais et al. 2002; Seager et al. 2012; Grenfell 2017; Kaltenegger 2017; Schwieterman et al. 2018). For such inhabited worlds to be positively identified from atmospheric spectra, they must possess global biospheres with a robust exchange of gases between life and the atmosphere as well as generate biosignature features that can be remotely detectable with foreseeable technologies and are separable from false-positive signals generated by abiotic processes (Catling et al. 2018; Meadows et al. 2018b; Schwieterman et al. 2018). Many groups have recently undertaken studies to assess the longevity and potential detectability of a variety of biosignature gases with near-future observational capabilities (Reinhard et al. 2017a; Fujii et al. 2018; Kawashima & Rugheimer 2019; Kaltenegger et al. 2020b; Pidhorodetska et al. 2020; Sousa-Silva et al. 2020; Checlair et al. 2021; Lin et al. 2021; Phillips et al. 2021; Wunderlich et al. 2021; Ranjan et al. 2022).

---
[14] https://exoplanets.nasa.gov/

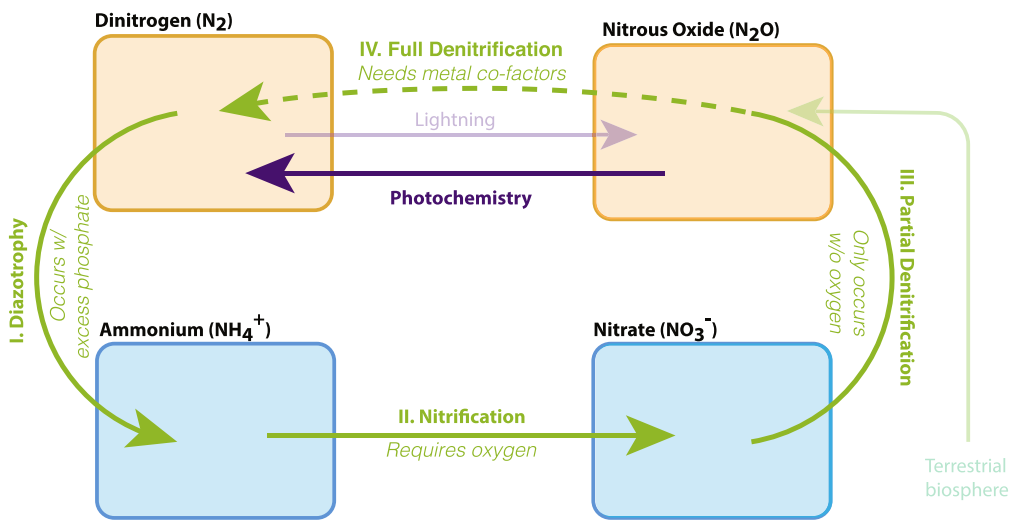

**Figure 1.** A schematic nitrogen cycle as implemented in the biogeochemical model cGENIE. The faded and dashed lines signify processes that are not explicitly included in the scheme. $N_2O$ fluxes result from incomplete denitrification of fixed nitrogen ($NO_3^-$) via the nitrous oxide reductase enzyme. Denitrification is an anaerobic process that depends on organic C fluxes, which are ultimately limited by nutrient ($PO_4^{3-}$) availability. A deficit of enzymatic catalysts—such as copper (Cu)—due to ocean water chemistry or initial planetary abundances would result in the partial termination of the N redox cycle and may produce large $N_2O$ fluxes. In the extreme scenario of no enzymatic catalysts, or if the nitrous oxide reductase enzyme (or an analog) had not evolved, the $N_2O$ flux would equal the total denitrification flux. Photochemical reactions will eventually return $N_2O$ to $N_2$ in the atmosphere.

The coexistence of $N_2$ and $O_2$ in an atmosphere is one possible biosignature. This possibility is due to both the high chemical disequilibrium between $N_2$, $O_2$, and surface liquid water (Krissansen-Totton et al. 2016, 2018) and the planetary evolution that is required to accumulate large quantities of both gases, including the implausibility of the abiotic generation and atmospheric retention of these gases together given competing atmospheric and geochemical sinks (Stüeken et al. 2016; Lammer et al. 2019; Sproß et al. 2021). However, $N_2$ is itself challenging to detect directly, due its status as a homonuclear molecule with no transitional dipole moment, possessing weak collisional-induced absorption bands at 4.3 and 2.15 $\mu$m (Lafferty et al. 1996; Schwieterman et al. 2015).

In addition to $N_2$ gas (in the atmosphere and dissolved in the ocean), nitrogen on the present-day Earth exists in a wide variety of different chemical forms, ranging from reduced $NH_4^+$ to oxidized $NO_3^-$, with many intermediate redox species in between, including $N_2O$. Overall, Earth's nitrogen cycle can be thought of as the biologically driven removal of $N_2$ from the ocean and atmosphere, the fixation of nitrogen in organic matter, which is then followed by the recycling of nitrogen back to the atmosphere as $N_2$ (Figure 1; Thamdrup 2012; Tian et al. 2015). However, as a reflection of the diversity of microbial metabolisms, the recycling loop contains multiple pathways back to $N_2$. One of these recycling routes involves the creation $N_2O$, which can either be further reduced biologically to $N_2$ or escape directly across the air–sea interface into the atmosphere.

Denitrification (the transformation of $NO_3^-$ to $N_2$ gas, with $N_2O$ as an intermediate product) is a relatively ubiquitous metabolism on Earth, and consequent $N_2O$ production can be mediated both by bacteria as well as by some fungi (Chen et al. 2015). Additionally, the direct oxidation of ammonia by bacteria and archaea can also produce $N_2O$ (Santoro et al. 2011; Prosser & Nicol 2012). Biological fluxes of $N_2O$ into the atmosphere are several orders of magnitude larger than abiotic sources, such as lightning, which is estimated to produce ∼0.002% of atmospheric $N_2O$ (Schumann & Huntrieser 2007). On Earth today, the magnitude of this biological flux is ∼0.4 Tmol yr$^{-1}$ and includes contributions from both marine and terrestrial (including agricultural and industrial) sources (Tian 2015; Tian et al. 2020).

$N_2O$ is of interest here because it produces notable features in Earth's near-IR (NIR) and MIR spectra (Sagan et al. 1993; Robinson & Reinhard 2018; Gordon et al. 2022). This, and its dominant biological origin on Earth, has led previous authors to consider $N_2O$ as a potential remote biosignature for Earthlike planets, along with $O_2$, $O_3$, and $CH_4$ (Rauer et al. 2011; Grenfell 2017; Schwieterman et al. 2018). In general, previous studies of $N_2O$ biosignatures have used present-day Earth's $N_2O$ flux as a fiducial boundary condition to predict the resulting mixing ratios on an Earthlike planet orbiting another star or have used an inferred $N_2O$ flux from Earth's geologic past as this surface boundary condition to predict mixing ratios to similar ends (Segura et al. 2003, 2005; Kaltenegger et al. 2007, 2020a; Rugheimer et al. 2013, 2015b; Grenfell et al. 2014; Tabataba-Vakili et al. 2016; Robinson & Reinhard 2018; Rugheimer & Kaltenegger 2018; Lin et al. 2021; Alei et al. 2022). Importantly, most of these past studies have typically predicted $N_2O$ mixing ratios on exo-Earths that are lower or only modestly higher than modern Earth, leading to pessimistic predictions for $N_2O$ detectability (e.g., Alei et al. 2022), though it has long been recognized that low stellar UV fluxes promote $N_2O$ accumulation (e.g., Segura et al. 2003, 2005; Grenfell et al. 2014; Rugheimer & Kaltenegger 2018) and that higher $N_2O$ surface fluxes would result in a more detectable biosignature (Kaltenegger 2017; Schwieterman et al. 2018).

It has been hypothesized that biological fluxes of $N_2O$ during the Proterozoic Eon (~2500–540 million years ago, Ma) may have been dramatically larger than at present, due to the limited availability of copper catalysts in euxinic (anoxic and sulfur-rich) oceans, which would have effectively short-circuited the last metabolic step in the denitrification cycle (Buick 2007; Figure 1). Higher atmospheric mixing ratios of $N_2O$ could therefore also have contributed to greenhouse warming at the time (Roberson et al. 2011), although lower $O_2$ concentrations would have somewhat muted the impact of higher fluxes, due to the reduced shielding of photolyzing UV radiation (Roberson et al. 2011; Stanton et al. 2018). Under certain conditions, $N_2O$ production on the earlier Earth could also have been augmented by chemodenitrification—that is, the process of $N_2O$ production via abiotic reduction of nitric oxide (NO) by ferrous iron (Samarkin et al. 2010; Stanton et al. 2018). Previous studies have not comprehensively examined the full range of plausible $N_2O$ fluxes over a range of $pO_2$ values and stellar types, including the end-member scenario, in which the nitrous oxide reductase enzyme that facilitates the last step in the denitrification process simply does not evolve (Pauleta et al. 2013). In such a scenario, we will show that $N_2O$ can accumulate to high concentrations—even for planets orbiting FGK stars—with implications for the detectability of this biosignature gas with current and upcoming observatories.

Here, we conduct a systematic photochemical and spectral investigation of $N_2O$ as an exoplanet biosignature and place upper limits on the $N_2O$ abundances and detectability from a productive biosphere. In Section 2, we use the Earth system (biogeochemical) model "cGENIE" to calculate denitrification fluxes for an Earthlike marine biosphere as a function of atmospheric oxygenation levels ($pO_2$) and concentrations of bioavailable phosphorous (as $PO_4^{3-}$). We evaluate our results against literature values for the Earth and generate realistic bounds for plausible intermediate and maximum $N_2O$ fluxes. In Section 3, we calculate the photochemical stability and steady-state mixing ratios of $N_2O$ given a large range of fluxes, inclusive of those calculated in Section 2, with a variety of oxygenation states and for stellar hosts that span the main sequence (F4V to M8V). In Section 4, we generate emission and transmission spectra for a subset of the scenarios investigated in Section 3. Finally, we calculate the number of transits of TRAPPIST-1e that will be required to detect $N_2O$ with JWST for three $N_2O$ flux scenarios, and find that detecting $N_2O$ with NIRSpec is plausible for production fluxes near the biospheric maxima. We discuss the implications and potential false positives in Section 5. We conclude in Section 6.

## 2. Circumscribing Plausible Global $N_2O$ Fluxes with a Biogeochemical Model

To map out how the total oceanic denitrification (and hence the potential maximum $N_2O$ production) rate varies as a function of $pO_2$ and phosphate availability ($PO_4^{3-}$), we use the cGENIE Earth system model of intermediate complexity. cGENIE consists of a 3D ocean circulation model plus a 2D energy balance and moisture model and a 2D sea-ice model. The 2D grid is split into $36 \times 36$ equal-area cells, while we adopt 16 depth layers in the ocean, following Cao et al. (2009). cGENIE simulates a 3D marine biosphere, including phosphorous and nitrogen-limited primary production, and a set of metabolisms, including aerobic respiration, anaerobic respiration, methanogenesis, and aerobic methanotrophy (Ridgwell et al. 2007; Olson et al. 2016). A simple 2D (not vertically resolved) gridded calculation of basic atmospheric chemical reactions is also included (Reinhard et al. 2020). cGENIE has been leveraged to explore the coupled evolution of Earth's biosphere, atmosphere, and climate system over the entire geologic timescale, including the Archean (e.g., Olson et al. 2013), Proterozoic (e.g., Olson et al. 2016; Reinhard et al. 2016, 2020), and Phanerozoic (e.g., Kirtland Turner & Ridgwell 2016). cGENIE has most recently been used to explore the relationship between planetary obliquity, nutrient cycling, and the consequent potential for atmospheric oxygenation on exoplanets (Barnett & Olson 2022).

The biological N cycle in cGENIE includes diazotrophy (the biological reduction of $N_2$ to $NH_4^+$, which can then be incorporated into biomass), nitrification (the oxidation of $NH_4^+$ to $NO_3^-$), and denitrification (the biological reduction of $NO_3^-$ to $N_2$). These processes are highlighted in Figure 1. Diazotrophy occurs only when N is scarce relative to phosphate ($PO_4^{3-}$) and N:P < 16 (the "Redfield Ratio") within the photic zone. Consequently, excess $PO_4^{3-}$ availability relative to N will drive greater diazotrophy, until the global rates of N fixation balance N loss. The rates of nitrification and denitrification are both sensitive to atmospheric $pO_2$, which directly influences surface and benthic oxygen concentrations, but their relationships to oxygen differ dramatically. Nitrification requires $O_2$, whereas denitrification occurs in the absence of $O_2$. Denitrification additionally requires reduced organic material and is a multistep process, with several intermediate N species between $NO_3^-$ and $N_2$, such as $N_2O$. The configuration of cGENIE employed here neglects this complexity and, by default, assumes the complete reduction of $NO_3^-$ to $N_2$ when organic matter and $NO_3^-$ are in sufficient abundance and local dissolved $O_2$ is low, following Naafs et al. (2019).

We estimate an upper bound on the possible $N_2O$ flux arising from incomplete denitrification for a given atmospheric $pO_2$ and ocean nutrient inventory by assuming that the entire denitrification flux results in the evolution of $N_2O$ (rather than going directly to $N_2$). This could occur, for instance, if the nitrous oxide reductase, the enzyme that facilitates the last step of the denitrification process (Pauleta et al. 2013), has not evolved or if dissolved copper, which is key to the functioning of this catalyst, was severely rate-limiting in abundance (Buick 2007). Nitrous oxide reductase can also be substantially inhibited in community settings by other biological products, including $C_2H_2$, CO, NO, $N_3^-$, and $CN^-$ (Kristjansson & Hollocher 1980; Koutný & Kučera 1999). Our goal here is not to be overly prescriptive of the specific scenarios in which denitrification is incomplete, but instead to examine plausible maxima in $N_2O$ production by Earthlike biospheres.

Figure 2 shows the total denitrification flux from Earth's marine biosphere as a function of $pO_2$ (relative to the present atmospheric level, or PAL) and phosphate availability ($PO_4^{3-}$, relative to the present ocean level, or POL). We simulate $pO_2$ levels of 0%–100% and phosphate availability between one and 2 times POL. The upper range of phosphate availability represents a planet with higher nutrient availability from enhanced continental weathering or a larger crustal abundance of P compared to Earth. Continental weathering could be enhanced by a more robust hydrological cycle, greater topographic relief, or a larger extent of coastal depositional settings. Indeed, enhanced weathering in the wake of snowball deglaciation may have resulted in $PO_4^{3-}$ concentrations transiently exceeding 2 times POL in the late Neoproterozoic, roughly coincident with the evidence for increases in oxygen levels (Planavsky et al. 2010). Steady-state $P$ levels are unlikely to have dramatically exceeded ∼2 times POL at any point in Earth's history (Reinhard et al. 2017b; Lenton et al. 2018; Reinhard & Planavsky 2022).

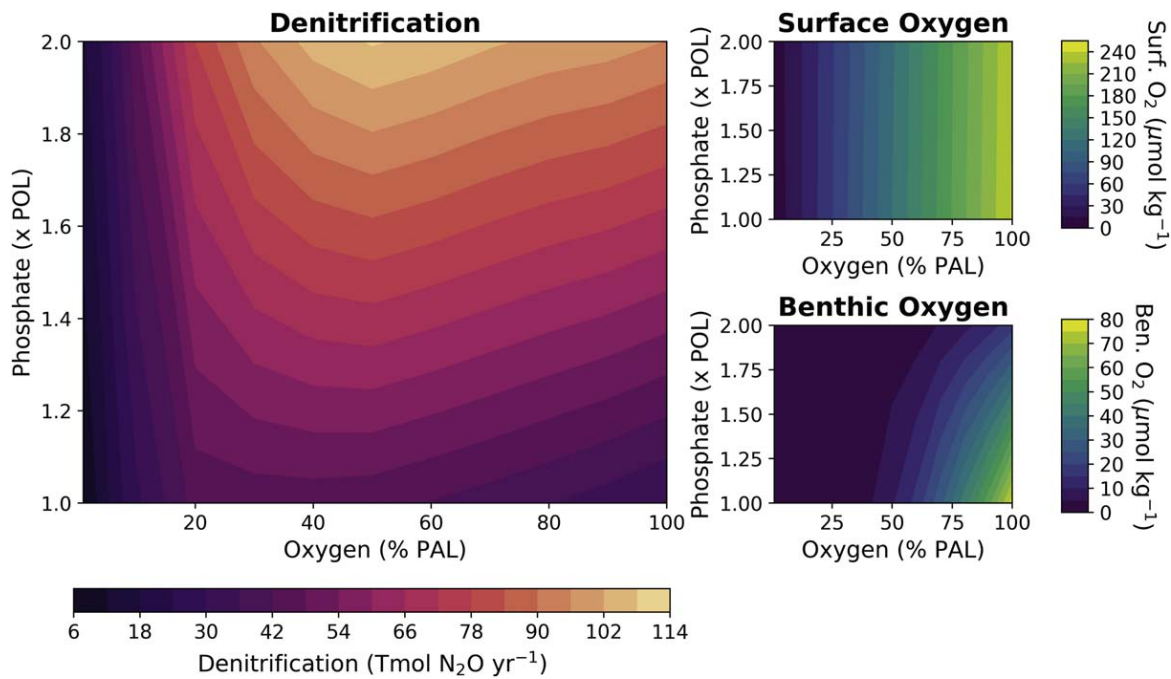

**Figure 2.** Results from GENIE showing (a) denitrification rates, (b) surface oxygen concentrations, and (c) benthic oxygen concentrations as a function of atmospheric oxygen (in terms of % of PAL; x-axis) and phosphate availability (in terms of POL; y-axis). Denitrification is optimized at intermediate atmospheric oxygen levels, where there is sufficient $O_2$ in the surface waters to stimulate surface nitrate production via nitrification, but insufficient $O_2$ to oxygenate the deep ocean, which would suppress denitrification.

The denitrification rates increase with P availability, which controls the organic C fluxes. The relationship between denitrification and atmospheric oxygen is more complex. At low levels of oxygen, nitrification, which requires oxygen, is limited. The denitrification rates are thus limited as well. At very high levels of oxygen, nitrification occurs readily, but denitrification, which requires low levels of oxygen, is suppressed. Oxygen is heterogeneously distributed in the ocean, such that both metabolisms may occur simultaneously, despite being spatially separated. Nitrification is favorable in well-oxygenated surface waters where oxygenic photosynthesis occurs, whereas denitrification is most favorable in oxygen minimum zones underlying productive regions of the surface ocean, where high organic C fluxes deplete $O_2$ via aerobic respiration. This possibility may be particularly true for atmospheric $pO_2$ that is lower than the present-day abundance.

Our results show a maximum denitrification flux of ∼40 Tmol yr$^{-1}$ (1 × POL P) to 100 + Tmol yr$^{-1}$ (2 × POL P) around an atmospheric $pO_2$ of ∼50% PAL $O_2$. At this intermediate oxygen level, the surface ocean is in equilibrium with the atmosphere and is well oxygenated, but the deep ocean remains poorly ventilated—optimizing the rates of both nitrification and denitrification. However, the sensitivity to $pO_2$ is asymmetric around this level. At the lowest $O_2$ levels (<20% PAL), denitrification is strongly attenuated (due to limited nitrification; see, e.g., Anbar & Knoll 2002; Fennel et al. 2005). Toward higher $O_2$ levels, denitrification falls off, with a shallower linear decrease, as deep ocean oxygenation increases. However, even under very high oxygen levels, such as those of modern Earth, oxygen minimum zones persist and allow denitrification in an otherwise well-oxygenated ocean.

To contextualize our calculations, the total denitrification flux on modern Earth is about ∼20 Tmol $N_2$ yr$^{-1}$, with substantial uncertainties (Canfield et al. 2010). As shown in Figure 2, a planet with twice the nutrient P availability could maintain a denitrification flux of ∼100 Tmol yr$^{-1}$ even at near-modern levels of $O_2$. This value is similar to a study of potential late Cretaceous (93 Ma) marine nitrogen cycling, for which Naafs et al. (2019) calculated a denitrification flux of ∼100 Tmol yr$^{-1}$, given 2 times POL, modern oxygen, and 4 times modern $CO_2$. We therefore adopt three fiducial $N_2O$ fluxes in our subsequent photochemical calculations of $N_2O$ abundances: 1, 10, and 100 Tmol yr$^{-1}$. A flux of 1 Tmol yr$^{-1}$ represents 5%–10% of the global denitrification flux of modern Earth having evolved as $N_2O$ rather than $N_2$, which is approximately a factor of 2 higher than Earth's estimated global $N_2O$ flux (Tian et al. 2020). A flux of 10 Tmol yr$^{-1}$ represents 50%–100% of the global denitrification flux of modern Earth (i.e., all $NO_3^-$ that is consumed in denitrification becomes $N_2O$), while a flux of 100 Tmol yr$^{-1}$ represents a global biosphere with lower oxygen and higher P than modern Earth, but consistent with periods of Earth's history. We regard the latter case as a reasonable upper bound for a weakly or fully oxygenated Earthlike world.

## 3. Calculating $N_2O$ Abundances for FGKM Stars with a Photochemical Model

Here we test the $N_2O$ flux–abundance photochemical relationships for a comprehensive range of $N_2O$ fluxes, which include the bounds described above (1, 10, and 100 Tmol yr$^{-1}$), but also extend to much lower fluxes. Note that we do not explicitly distinguish fluxes from the ocean and fluxes from a terrestrial biosphere in our atmospheric calculations. For comparison, the total primary production of the land-based denitrification is about one half that of the ocean (with substantial uncertainty; see, for example, Falkowski et al. 2000; Gruber & Galloway 2008). This difference is relatively small in comparison to the increase in denitrification when increasing the oceanic P availability from 1 to 2× POL, and we thus consider plausible terrestrial fluxes to be broadly included within these original bounds.

### 3.1. Photochemical Model and Inputs

To calculate flux–abundance relationships for Earthlike planets as a function of $N_2O$ flux and $pO_2$, we use the photochemical model component of the Atmos code[15] (Arney et al. 2016). The code was originally developed by Kasting et al. (1979) and has been improved by successive authors (Pavlov et al. 2001; Zahnle et al. 2006; Arney et al. 2016; Lincowski et al. 2018). The photochemical code uses the reverse Euler method to solve the flux and continuity equations at each vertical layer, providing stable solutions at steady state. The model uses a $\delta$ two-stream method to calculate the radiative transfer (Toon et al. 1989) and includes vertical transport via molecular and eddy diffusion. The atmosphere is divided into 200 layers of 0.5 km in altitude. The model contains NO production by lightning (Harman et al. 2018) and the $H_2O$ cross-sectional and sulfur gas reaction rate updates recommended by Ranjan et al. (2020).

To enhance reproducibility, we use the publicly available Atmos "ModernEarthSimple" template. This template includes 50 species and 238 photochemical reactions and is appropriate for modeling major trace species ($O_3$, $CH_4$, CO, and $N_2O$) on high-oxygen Earthlike planets (e.g., Meadows et al. 2018a). Table 1 contains our assumed surface boundary conditions, including the deposition rates and volcanic fluxes. These boundary conditions are consistent overall with those of the modern Earth and those used in previous studies of $O_2$-rich planets (e.g., Segura et al. 2005; Schwieterman et al. 2019a, 2019b; Wunderlich et al. 2020). We assume a variety of $O_2$ abundances, ranging from 0.01–1.0 PAL, which is equivalent to 0.002 to 0.21 bar. Our $N_2O$ surface fluxes range from 0.01 to 100 Tmol yr$^{-1}$ ($3.7 \times 10^7$ to $3.7 \times 10^{11}$ molecules cm$^{-2}$ s$^{-1}$). To further enhance reproducibility and isolate the sensitivity to varied molecular fluxes and stellar spectra, we assume a surface pressure of $P_0 = 1$ bar and a surface temperature of 288 K for all cases with Earth's modern temperature–pressure profile. $N_2$ is used as a filler gas. We compared our results to those obtained with the "ModernEarthComplex" template (based on Lincowski et al. 2018), which includes 71 additional reactions (309 total reactions) and 23 additional (73 total) chemical species, and found the predicted $N_2O$ mixing ratios to be consistent between templates. We also examined the sensitivity to stratospheric

---

[15] https://github.com/VirtualPlanetaryLaboratory/atmos

**Table 1**
Photochemical Boundary Conditions

| Chemical Species | Deposition Velocity (cm s$^{-1}$) | Flux (Molecules cm$^{-2}$ s$^{-1}$) | Mixing Ratio |
|---|---|---|---|
| O | 1 | … | … |
| $O_2$ | … | … | Variable |
| $N_2$ | … | … | Variable |
| $CO_2$ | $5 \times 10^{-5}$ | $6.9 \times 10^8$ | … |
| $H_2O$ | … | … | Fixed[a] |
| H | 1 | … | … |
| OH | 1 | … | … |
| $HO_2$ | 1 | … | … |
| $H_2O_2$ | 0.2 | … | … |
| $H_2$ | $2.4 \times 10^{-4}$ | … | … |
| CO | $1.2 \times 10^{-4}$ | $3.0 \times 10^{11}$ | … |
| HCO | 1 | … | … |
| $H_2CO$ | 0.2 | … | … |
| $CH_4$ | 0 | $1 \times 10^{11}$ | … |
| $CH_3$ | 1 | … | … |
| NO | $1.6 \times 10^{-2}$ | $1 \times 10^9$ | … |
| $NO_2$ | $3 \times 10^{-3}$ | … | … |
| HNO | 1 | … | … |
| $H_2S$ | 0.2 | $2 \times 10^8$ | … |
| $SO_2$ | 1 | $9 \times 10^9$ | … |
| $H_2SO_4$ | 1 | $7 \times 10^8$ | … |
| HSO | 1 | … | … |
| $O_3$ | 0.07 | … | … |
| $HNO_3$ | 0.2 | … | … |
| $N_2O$ | … | Variable | … |
| $HO_2NO_2$ | 0.2 | … | … |
| OCS | 0.01 | $1.57 \times 10^7$ | … |

**Notes.**
[a] The tropospheric $H_2O$ profile is fixed to an Earth average (Manabe & Wetherald 1967).
[b] The species included in the photochemical scheme with a deposition velocity and flux of 0 include $C_2H_6$, HS, S, SO, $S_2$, $S_4$, $S_8$, $SO_3$, $S_3$, N, $NO_3$, and $N_2O_5$.

temperature profiles and found minimal differences that are small compared to our range of considered $pO_2$ levels, $N_2O$ fluxes, and stellar spectra. Tropospheric temperature profiles should be relatively unaffected, assuming the same surface temperature of 288 K. Substantially different surface temperatures would impact $H_2O$ abundances (depending on the relative humidity), which can have downstream impacts on $CH_4$ and other trace gases, but will have smaller influences on $N_2O$, given its major photochemical sinks (see below).

We sourced stellar spectra directly from the existing Atmos library, including the additions from Arney (2019). Figure 3 shows the stellar spectra used in our simulations, with the bottom panel zooming in on the UV component of each spectrum and the molecular cross sections for $N_2O$, $O_2$, $O_3$, and $CH_4$. Our solar spectrum was sourced from Thuillier et al. (2004). The original source of the spectrum for the star HD 85512 (K6V) is the Measurements of the Ultraviolet Spectral Characteristics of Low-mass Exoplanetary Systems treasury survey (Youngblood et al. 2016; France et al. 2016; Parke Loyd et al. 2018), and the original source for the Proxima Centauri (M5V) spectrum is the Habitable Zones and M dwarf Activity across Time program (Shkolnik & Barman 2014; Parke Loyd et al. 2018; Peacock et al. 2020). The TRAPPIST-1 spectrum was the median average from the three-activity models simulated by Peacock et al. (2019a, 2019b). Note that for TRAPPIST-1 and Proxima Centauri specifically, we adopted flux scaling consistent with TRAPPIST-1e and

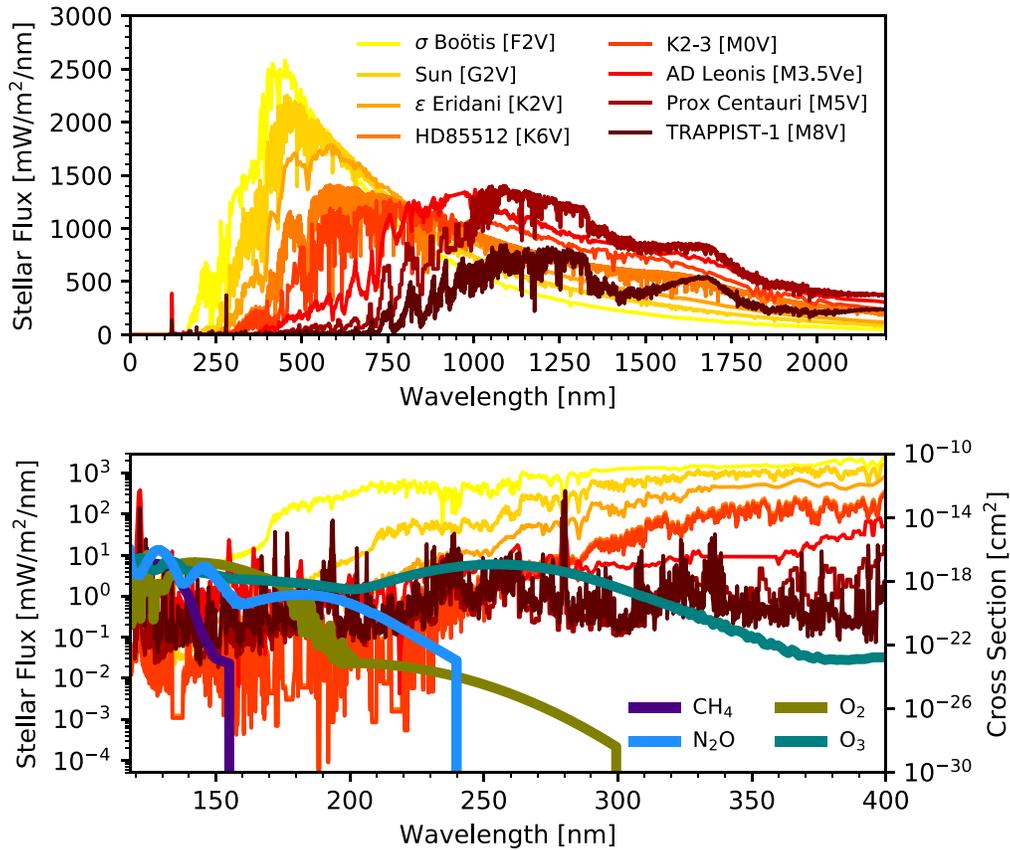

**Figure 3.** The top panel shows the stellar spectra used in our simulations. The bottom panel shows the UV wavelengths of the stellar spectra and the UV cross sections of $N_2O$, $O_2$, $O_3$, and $CH_4$ that are used in our simulations (the right $y$-axis).

Proxima Centauri b, while the other stars are scaled such that the simulated Earth-twin planet lies at the same Earth–Sun adjusted insolation value. However, the simulations using these scaling relationships are not meant to represent those planets precisely, since the surface gravity is not adjusted, and we do not account for potential synchronous rotation, to facilitate clear intercomparison of all photochemical results. For the case of the simulated transit spectra of TRAPPIST-1e, we adopted the planetary parameters, including surface gravity, that are described in Agol et al. (2021), and recalculated the photochemical results based on these values.

### 3.2. Predicted $N_2O$ Flux–Abundance Relationships for FGKM Stars

Figures 4 and 5 show comprehensive flux–abundance relationships for $N_2O$ as a function of the $N_2O$ surface molecular flux and $pO_2$, using the modeling inputs described in Section 3.1 for Earthlike planets orbiting stars with spectral types F4V to M8V: Sigma Boötis (F4V), the Sun (G2V), Epsilon Eridani (K2V), HD 85112 (K6V), K2-3 (M0V), AD Leonis (M3.5V), Proxima Centauri (M5V), and TRAPPIST-1e (M8V). In addition to the colored shading that indicates the calculated $N_2O$ concentrations, line contours are added to indicate the 0.1, 1.0, 10, 100, and 1000 ppm $N_2O$ levels. Figure 6 shows the flux–abundance relationships for the 100% PAL (21% v/v $O_2$) cross section of this data for all stars. Figures 7 and 8 display a subset of this data in an alternative visualization format for $pO_2 = 100\%$ PAL (Figures 7) and $pO_2 = 50\%$ PAL (Figure 8). (We also provide the $pO_2 = 10\%$ PAL case in Appendix Figure A1).

From these photochemical flux–abundance simulations, we can observe that for a given star, the $N_2O$ abundance is strongly dependent on the flux and only modestly dependent on the $O_2$ abundance (though we caution that the $N_2O$ flux can be highly sensitive to $pO_2$, as shown in Figure 2). The predicted $N_2O$ concentration is also very sensitive to the spectral type of the host star, with the largest enhancement seen for the late K dwarf HD 85112 (K6V) and the earliest M dwarf K2-3 (M0V). In comparison, $\sigma$ Boötis (F4V) and the Sun (G2V) show the lowest $N_2O$ buildup at any given $N_2O$ surface flux (with $\sigma$ Boötis being very slightly lower), while the mid-to-late M-dwarf planets maintain considerably higher $N_2O$ concentrations than the Sun-like case at any given flux, but nonetheless maintain lower $N_2O$ concentrations than the K-dwarf planets. The relationship between $N_2O$ and $O_2$ is strongest for the F and G dwarfs, weaker for the M dwarfs, and weakest for the K6V and M0V stars, which maintain comparatively high $N_2O$ at high fluxes, even for low $O_2$ levels.

These flux–abundance relationships result from the relative balance of photolysis and the other photochemical reactions that determine the atmospheric lifetime of $N_2O$. The major photochemical sink for $N_2O$ is photolysis ($N_2O + h\nu$ [$\lambda < 240$ nm] $\rightarrow N_2 + O(^1D)$), where photolysis proceeds most strongly via photons with wavelengths of less than 200 nm but weaker photolysis cross sections of up to 240 nm. Importantly, the wavelength-dependent overlap of the $N_2O$ cross sections between different stellar hosts drives large differences in the estimated photolysis rates and, consequently, the photochemical lifetimes and estimated abundances (Figure 3). At a fixed surface molecular flux of $N_2O$, the trend of increasing $N_2O$ abundances as a function of $pO_2$ results from the shielding impact of the Schumann–Runge O$_2$ bands (175–195 nm) and, to a lesser extent, the ozone (O$_3$) Hartley band at longer wavelengths (200–310 nm). Larger oxygen concentrations result in greater O$_2$ shielding, enhancing N$_2$O lifetimes and abundances. Because this shielding is more directly reliant on O$_2$ shielding than O$_3$ (which is logarithmically dependent on O$_2$), the relationship between N$_2$O abundances and O$_2$ is smooth and largely linear in log–log space.

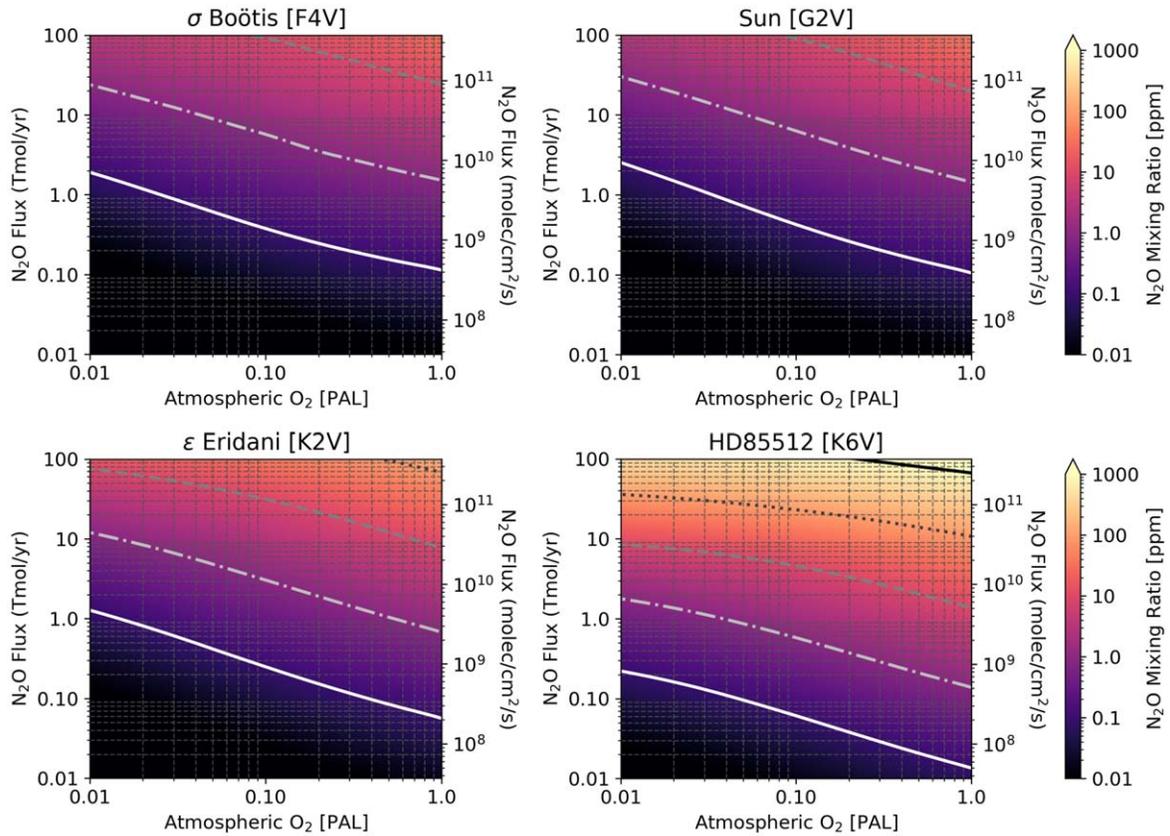

**Figure 4.** The atmospheric N$_2$O volume-mixing ratio (in ppm) as a function of atmospheric oxygen level in terms of PAL (*x*-axis) and the N$_2$O flux (*y*-axes). Each panel shows the results for an Earth analog at the inner edge of the habitable zone for Sigma Boötis [F4V], the Sun [G2V], Epsilon Eridani [K2V], and HD 85112 [K6V]. The solid white, light gray dotted–dashed, dark gray dashed, black dotted, and solid black lines represent 0.1, 1.0, 10, 100, and 1000 ppm N$_2$O contours, respectively. The surface boundary conditions are given in Table 1.

A consequential secondary photochemical sink for N$_2$O is the interaction with the O($^1$D) radical that can proceed through two channels: (1) N$_2$O + O($^1$D) → N$_2$ + O$_2$ and (2) N$_2$O + O($^1$D) → NO + NO. The O($^1$D) that reacts with N$_2$O can be sourced from N$_2$O photolysis, but for Sun-like host stars the substantially greater source is photolysis of the tropospheric ozone, via the reaction O$_3$ + h$\nu$ ($\lambda$ < 330 nm) → O($^1$D) + O$_2$. This reaction channel is strongly dependent on a relatively narrow range of near-UV (NUV) photons that have sufficient energy to photolyze O$_3$, but are low enough in energy to avoid being absorbed by overlying O$_2$ and O$_3$ in the stratosphere before reaching the troposphere. These NUV photons are produced by the stellar photosphere, and so their flux is strongly dependent on the effective temperature of the host star (see Figure 3). Earthlike planets orbiting late-type stars are particularly poor generators of atmospheric O($^1$D) radicals (Segura et al. 2005; Grenfell et al. 2013, 2014; Rugheimer et al. 2015a; Arney 2019), and so this reaction channel is correspondingly weaker for K-host stars versus G-host stars, and weakest for M-dwarf stars. Our model also includes a rainout sink for N$_2$O (Lincowski et al. 2018), which is a tertiary sink not dependent on stellar type.

The balance between the photolysis and O($^1$D) destruction channels is shown as a function of the surface N$_2$O mixing ratio for a subset of the stars examined in Appendix Figure A2. For all stars, photolysis causes the greatest loss rate for N$_2$O. For Sun-like G and F stars, reactions with O($^1$D) radicals are also a robust, though secondary, sink for N$_2$O. For the late K-dwarf stars, the O($^1$D) sink is intermediate between those for the G and M dwarfs, but the photolysis rate is the lowest among all possibilities. For M-dwarf stars, the O($^1$D) sink for N$_2$O is weak (Grenfell et al. 2013, 2014), but increased magnetic activity enhances the far-UV (FUV) radiation that can photolyze N$_2$O over relatively inactive K-dwarf stars. These photochemical loss relationships combine to put the latest K-dwarf hosts in the "sweet spot," with the lowest integrated N$_2$O loss rates over both types of photochemical sink, and hence the highest predicted concentrations at any fixed N$_2$O surface flux and $pO_2$ parameter combination.

We find that the FUV continuum has a crucial impact on the M-dwarf N$_2$O photolysis rates; it is only very weakly a function of the Ly$\alpha$ flux (Grenfell et al. 2014; Peacock et al. 2022), so the scaling or estimation of stellar spectra that do not account for the FUV continuum will yield divergent results from those presented here. This finding is consistent with the recent work of Teal et al. (2022), who focused on the impact of differences in the input stellar spectra on the signatures of CH$_4$ and haze on Archean-like planets. We note that models that artificially remove or substantially scale down the FUV continuum for M dwarfs (i.e., "inactive" stellar models) will instead show that inactive late M dwarfs, rather than late K dwarfs, produce the largest $N_2O$ concentrations for a given flux, due to strongly attenuated $N_2O$ photolysis (e.g., Grenfell et al. 2014; Rugheimer et al. 2015a). While modeling inactive M-dwarf photochemistry provides an informative sensitivity analysis, it is not representative of observed and modeled mid-to-late M-dwarf spectra, which show substantial FUV continua (e.g., Peacock et al. 2019a, 2019b).

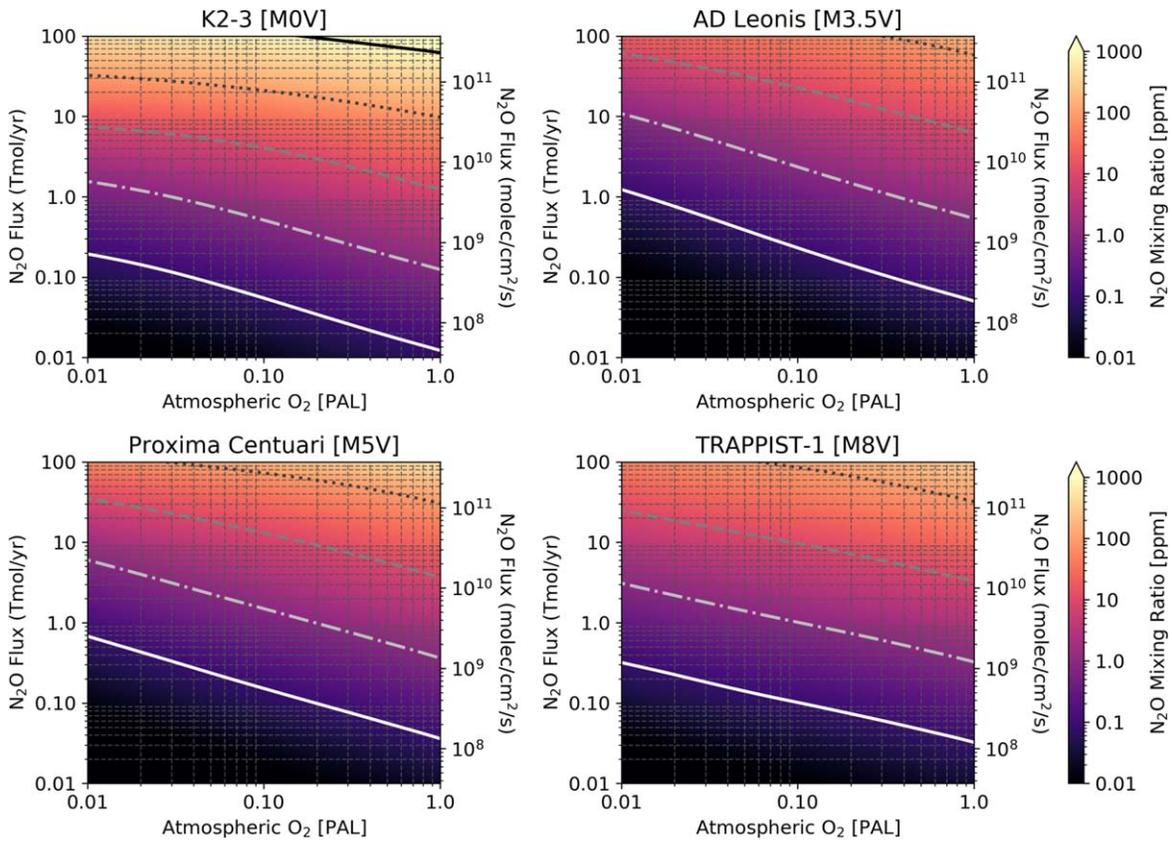

**Figure 5.** The same as Figure 4, but results are shown for Earthlike planets at the inner edge of the habitable zone of the M-dwarf stars K2-3 [M0V], AD Leonis [M3.5V], Proxima Centauri [M5V], and TRAPPIST-1e [M8V].

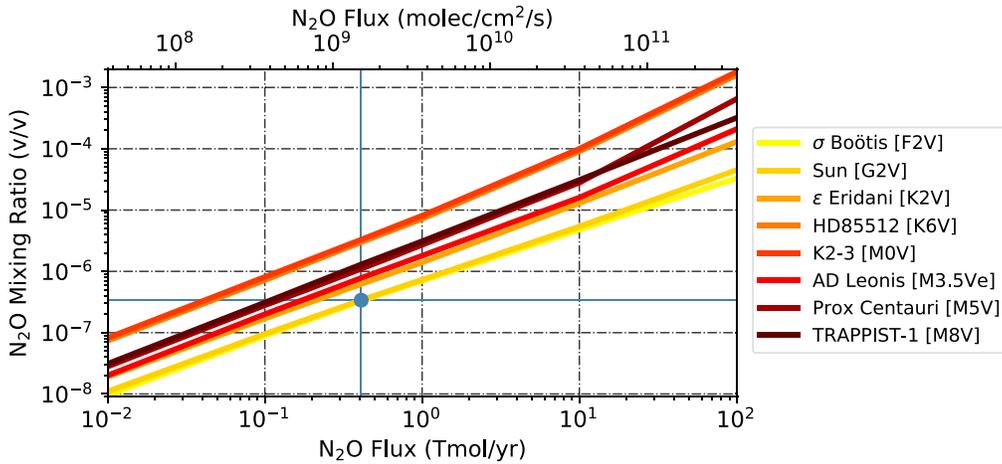

**Figure 6.** Atmospheric mixing ratios of $N_2O$ as a function of the surface fluxes for an Earthlike atmosphere with $pO_2 = 100\%$ PAL. Results are shown for a range of stellar host stars (FGKM) and represent a vertical slice through the rightmost axis of the panels shown in Figures 4 and 5. The blue horizontal line shows the mixing ratios for $N_2O$ in the modern atmosphere (∼330 ppmv), while the blue vertical line indicates the modern $N_2O$ surface flux (∼0.4 Tmol yr$^{-1}$ or $1.5 \times 10^9$ molecules cm$^{-2}$ s$^{-1}$).

The flux–abundance relationships allow an estimation of the envelope of plausible $N_2O$ concentrations given more or less efficient denitrification in exo-biospheres, where $N_2O$ is evolved rather than $N_2$. For an Earth orbiting a solar twin (G2V), we find that $N_2O$ fluxes of 10–100 Tmol yr$^{-1}$ would lead to maximum $N_2O$ abundances of 5–50 ppm for Earth–Sun analogs, compared to $\sim 330$ ppb for modern Earth, given a surface $N_2O$ molecular flux of $\sim 0.4$ Tmol yr$^{-1}$. The F4V dwarf stellar host $\sigma$ Boötis tracks very closely to the Sun, with slightly lower $N_2O$ mixing ratios for a given flux (see Figures 4 and 6). For the quiescent late K dwarf HD 85512 (K6V) and earliest M dwarf K2-3 (M0V), $N_2O$ fluxes of 10–100 Tmol yr$^{-1}$ would result in $N_2O$ mixing ratios of 90–1600 ppm. This is important, because $N_2O$ concentrations are increased by about two orders of magnitude for a given flux for an exo-Earth orbiting a late K-dwarf star versus a solar analog. Indeed, the modern $N_2O$ flux on Earth would result in $N_2O$ concentrations of $\sim 3$ ppm for Earth analogs orbiting these host stars, compared to the 330 ppb modern $N_2O$ concentration. We note that these $N_2O$ concentration envelopes are most applicable to planets with Earthlike insolations and Earthlike $CO_2$ values. Habitable planets that are located in the outer habitable zone will experience lower UV fluxes, reducing the major sink of $N_2O$, and gain additional $CO_2$ shielding, assuming that $CO_2$ is the major greenhouse gas.

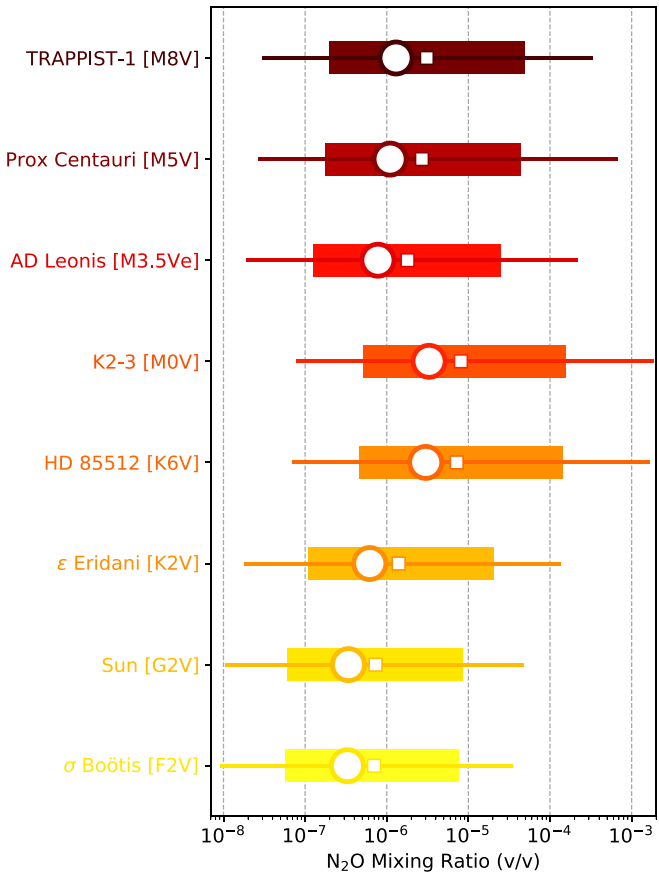

**Figure 7.** Steady-state ground-level mixing ratios of $N_2O$ for modern Earthlike planets around a range of stellar hosts for an Earthlike planet with $pO_2 = 100\%$ PAL. The open circles show the results for a modern $N_2O$ flux of $\sim 0.4$ Tmol yr$^{-1}$ ($1.5 \times 10^9$ molecules cm$^{-2}$ s$^{-1}$), the open squares show the results for an $N_2O$ flux of 1 Tmol yr$^{-1}$, the shaded bars show the range of 0.1–10 Tmol yr$^{-1}$ (a factor of 10), and the horizontal lines show a range of 0.01–100 Tmol yr$^{-1}$ (a factor of 100). A flux of 1 Tmol yr$^{-1}$ corresponds to a surface molecular flux of $3.7 \times 10^9$ molecules cm$^{-2}$ s$^{-1}$.

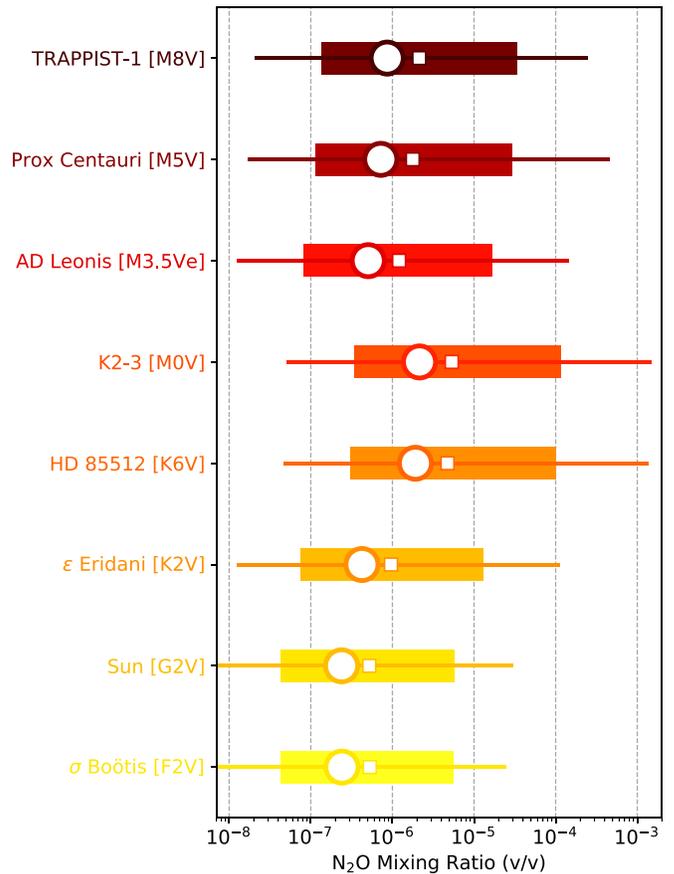

**Figure 8.** The same as Figure 7, but for an Earthlike planet with $pO_2 = 50\%$ PAL.

We tested the sensitivity of our results to different $N_2$ partial pressures (Figure A3) and increases in assumed eddy diffusion (Figure A4). In general, the trends described here are robust to large changes in these variables. We direct the reader to the Appendix for more detailed descriptions of these sensitivity experiments.

### 3.3. Atmospheric Profiles for Spectral Simulations

We refined a subset of our flux–abundance calculations to prepare self-consistent chemical profiles as inputs for spectral simulations. We specifically consider an Earth–Sun scenario, an Earth orbiting HD 851512 (K6V), an Earthlike Proxima Centauri b, and an Earthlike TRAPPIST-1e. We adjust the surface gravities of Proxima Centauri b and TRAPPIST-1e so that they are consistent with values reported in the literature (Agol et al. 2021; Faria et al. 2022). Otherwise, we assume Earth's surface gravity (9.8 m s$^{-2}$). For all scenarios, we calculate chemical profiles based on 1, 10, and 100 Tmol yr$^{-1}$ $N_2O$ surface fluxes. We continue to assume an Earthlike surface temperature of 288 K and a surface pressure of 1 bar. Figure 9 shows the resulting altitude-dependent chemical profiles for select trace gases ($N_2O$, $O_3$, $CH_4$, and CO). The $N_2O$ profiles increase sensibly with flux, while the other trace gases, including $CH_4$, CO, and $O_3$, decrease with increasing $N_2O$ flux. These decreases in trace gas abundance are primarily due to the increase in $O(^1D)$ radicals liberated by $N_2O$ photolysis ($N_2O + h\nu$ [$\lambda < 240$ nm] $\rightarrow$ $N_2 + O(^1D)$). The $O(^1D)$ radicals directly or indirectly (e.g., via the downstream production of OH) attack trace gases, including $CH_4$, CO, and $O_3$. The $O_3$ is additionally destroyed via cycles catalyzed by NO, which is liberated from $N_2O$ photolysis, and is thus strongly dependent on the $N_2O$ abundance (Ravishankara et al. 2009). The magnitude of the impact of $N_2O$ on other trace gases is inversely related to the photospheric temperature of the host star, and is particularly muted in the TRAPPIST-1e case. Note that while increasing the $N_2O$ fluxes leads to less abundant $CH_4$, the overall $CH_4$ mixing ratios are considerably higher for the M-dwarf planets than for the G- or K-dwarf planets, as expected (Segura et al. 2005; Rugheimer et al. 2015a).

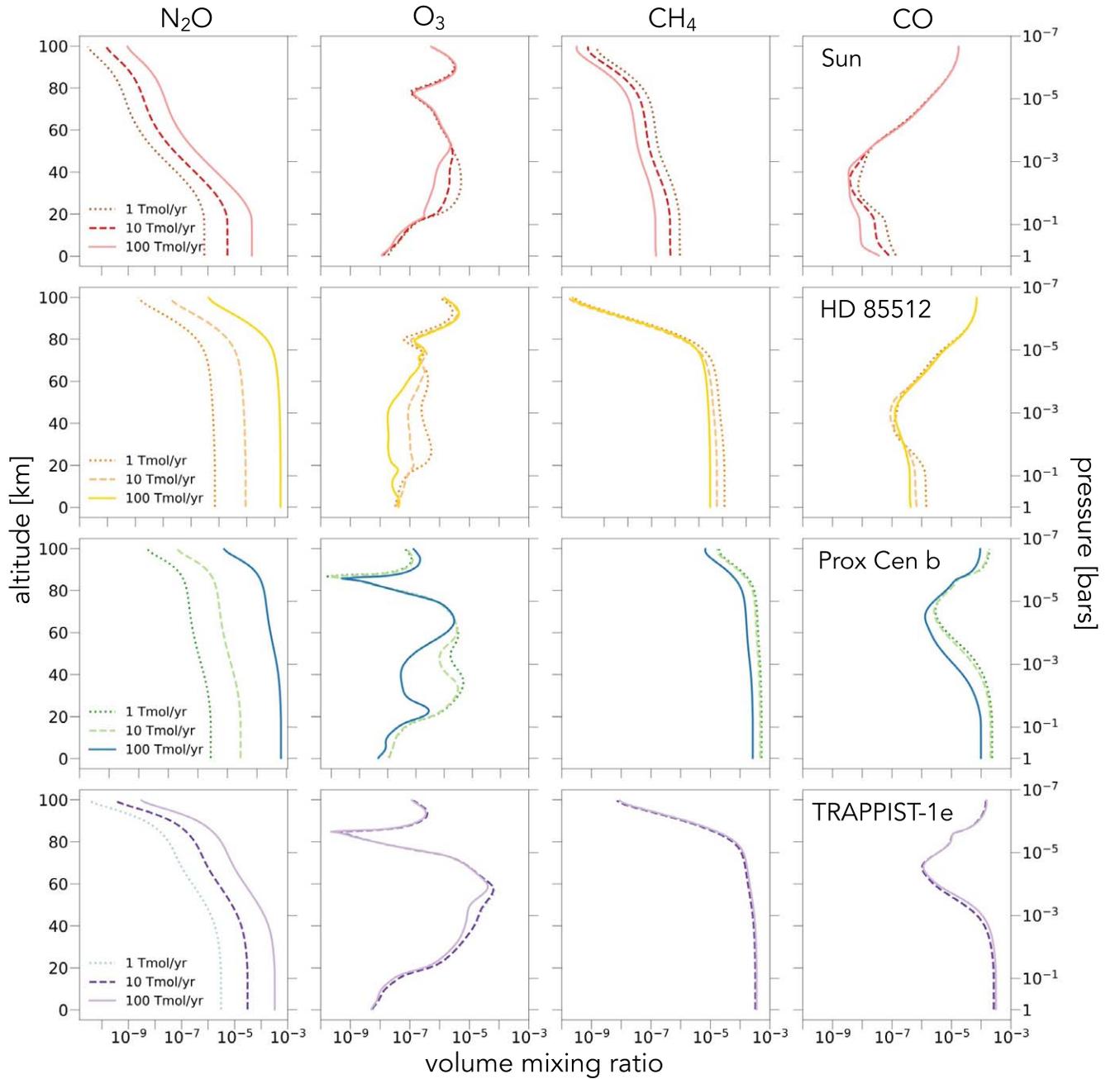

**Figure 9.** Gas mixing ratio profiles for (from top to bottom): the Earth orbiting the Sun, an Earthlike planet orbiting the late K-dwarf star HD85512 (K6V), an Earthlike Proxima Centauri b, and an Earthlike TRAPPIST-1e. The $N_2O$ profiles assuming fluxes of 1, 10, and 100 Tmol yr$^{-1}$ are shown. For brevity, not all species are included. All runs assume $pO_2 = 100\%$ PAL.

## 4. $N_2O$ Observables in the NIR and MIR

### 4.1. Spectral Simulation Tool

We use the Planetary Spectrum Generator (PSG) to simulate emission and transmission spectra of select cases that are explored in Section 3. PSG is a versatile and publicly accessible radiative transfer tool that is used to simulate remote observables for a wide variety of planetary objects and viewing geometries, which can calculate synthetic noise for a variety of instrumental configurations (Villanueva et al. 2018, 2022). The input IR opacities of PSG are sourced from the HITRAN database (Gordon et al. 2022). PSG has been widely used for the forward modeling of planetary spectra and the calculation of anticipated noise sources, particularly the cases of terrestrial exoplanetary atmospheres (Fauchez et al. 2020; Suissa et al. 2020; Pidhorodetska et al. 2020, 2021).

### 4.2. Simulated Thermal Emission Spectra

We used PSG to calculate synthetic thermal emission spectra from 5 to 20 $\mu$m for planets with the chemical profiles shown in

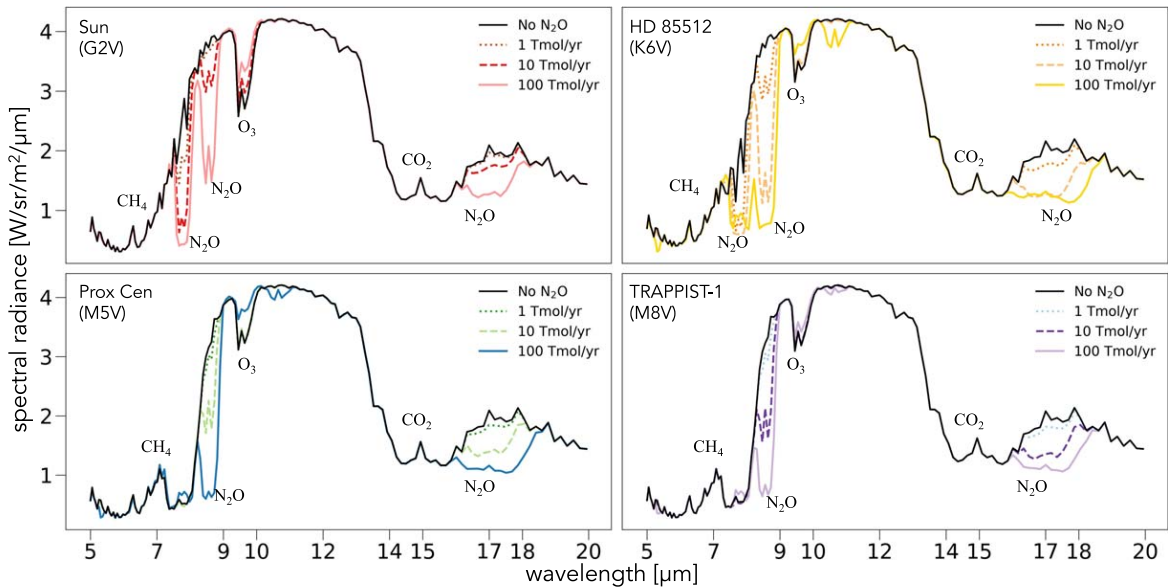

**Figure 10.** Simulated thermal emission spectra of (a) an Earth–Sun scenario, (b) an Earth orbiting the late K-dwarf star HD85512 (K6V), (c) an Earthlike Proxima Centauri b, and (d) an Earthlike TRAPPIST-1e. Each figure shows variations in the spectra according to $N_2O$ surface fluxes of 1, 10, and 100 Tmol yr$^{-1}$, using the temperature and gas mixing ratio profiles shown in Figure 9 and assuming 50% cloud cover.

Figure 9. We assumed 50% cloud cover for all cases, with half liquid water cloud particles and half cirrus cloud particles. Planetary thermal emission spectra could be targeted by future space-based interferometers, such as the LIFE mission concept (Quanz et al. 2018, 2019, 2022; Defrère et al. 2018; Alei et al. 2022). Direct imaging in the thermal IR is also a possibility for a handful of systems with ground-based 30 m class telescopes (Fujii et al. 2018; López-Morales et al. 2019). We did not attempt to quantify the relative detectability of the features modeled in emission, since the parameters and capabilities of these future observations are not yet well defined. However, we note that these simulations will be useful inputs for future ground- and space-based thermal IR detectability studies.

Figure 10 shows indicative planetary thermal IR spectra of four scenarios: (1) an exo-Earth orbiting a solar twin; (2) an exo-Earth orbiting the K6V star HD 85512; (3) an Earthlike Proxima Centauri b; and (4) an Earthlike TRAPPIST-1e. These scenarios are hypothetical in nature and likely do not represent the actual planets, but are self-consistent representations of modern $O_2$-rich Earthlike worlds orbiting stars of similar spectral type. Each scenario shows the resulting spectra for no $N_2O$ flux and for $N_2O$ surface fluxes of 1, 10, and 100 Tmol yr$^{-1}$. The major $N_2O$ bands in the thermal IR have band centers at 7.8, 8.5, and 17 $\mu$m. Among these, the 8.5 $\mu$m band has the highest intrinsic opacity (Gordon et al. 2022). As anticipated, the depths of these absorption bands are a strong function of surface flux and the host star spectrum. The strongest $N_2O$ bands at a given flux are seen for the K6V host, while those for the hypothetical Earthlike Proxima Centauri b and TRAPPIST-1e are markedly greater at each given flux than those for the Earth–Sun scenario.

For the exo-Earth orbiting the K6V host, $N_2O$ absorption bands become comparable to other biosignature absorption features, such as $O_3$ at 9.65 $\mu$m, at an $N_2O$ surface flux of only 1 Tmol yr$^{-1}$ (around 2.5 times the modern Earth's globally averaged $N_2O$ flux). For $N_2O$ surface fluxes of 1–10 Tmol yr$^{-1}$, $N_2O$ bands become comparable to $O_3$ for the Proxima Centauri b and TRAPPIST-1e cases. For the Earth–Sun scenario, $N_2O$ surface fluxes must be above 10 Tmol yr$^{-1}$ for the absorption depth of the 8.5 $\mu$m $N_2O$ band to become comparable to the 9.65 $\mu$m $O_3$ band. At 100 Tmol yr$^{-1}$, an additional but intrinsically weak band of $N_2O$ becomes noticeable for the K6V case centered at 10.6 $\mu$m. This may be a particularly useful diagnostic of extremely high $N_2O$ fluxes, comparable to the anticipated denitrification flux of the entire planetary biosphere.

From the spectra plotted in Figure 10, interactions between the detectability of biosignature gases become notable. The depth of the 9.65 $\mu$m $O_3$ band is strongly influenced by the $N_2O$ flux for the Earth–Sun case, which has somewhat less of an effect for the K6V and M5V scenarios, and a minimal effect for the TRAPPIST-1 case. These observations are consistent with the depletions in chemical profiles calculated as inputs in Figure 9, as discussed in Section 3.3. Importantly, the 7.8 $\mu$m $N_2O$ band is obscured in the Proxima Centauri b and TRAPPIST-1e atmospheres by the 7.7 $\mu$m $CH_4$ band. The methane is substantially more abundant in these atmospheres at a given flux than in the G- or K-dwarf planets; however, the $N_2O$ abundance could still be estimated from the 8.5 $\mu$m $N_2O$ band, which is not substantially contaminated by $CH_4$ absorption.

We emphasize that the MIR spectral results presented here differ from previous work primarily because we consider the potential for larger biogenic $N_2O$ fluxes, informed by our biogeochemical model, rather than a fundamentally different treatment of photochemistry or spectral simulation. For example, Rugheimer & Kaltenegger (2018) predicted biosignature abundances, including $N_2O$, based on biological fluxes inferred from Earth–Sun mixing ratios, including those estimated for Earth's earlier geologic epochs. The $N_2O$ mixing ratios for earlier Earth eons were estimated to be lower than those of the present day (∼0–100 ppb), because of less efficient UV shielding due to lower atmospheric oxygen concentrations, but with biogenic $N_2O$ fluxes that are similar to those of Earth today. Consequently, Rugheimer & Kaltenegger (2018) find that, given these assumptions, $N_2O$ does not contribute notably to the emitted light spectrum, and would likely be undetectable in a low-resolution and photon-limited exoplanet spectrum for each case that they investigated. We note that those cases assume an N$_2$O signature that is smaller than those we estimate for the 1 Tmol yr$^{-1}$ case presented in Figure 10, and so our overall results are comparable when comparing equivalent N$_2$O fluxes and oxygen concentrations.

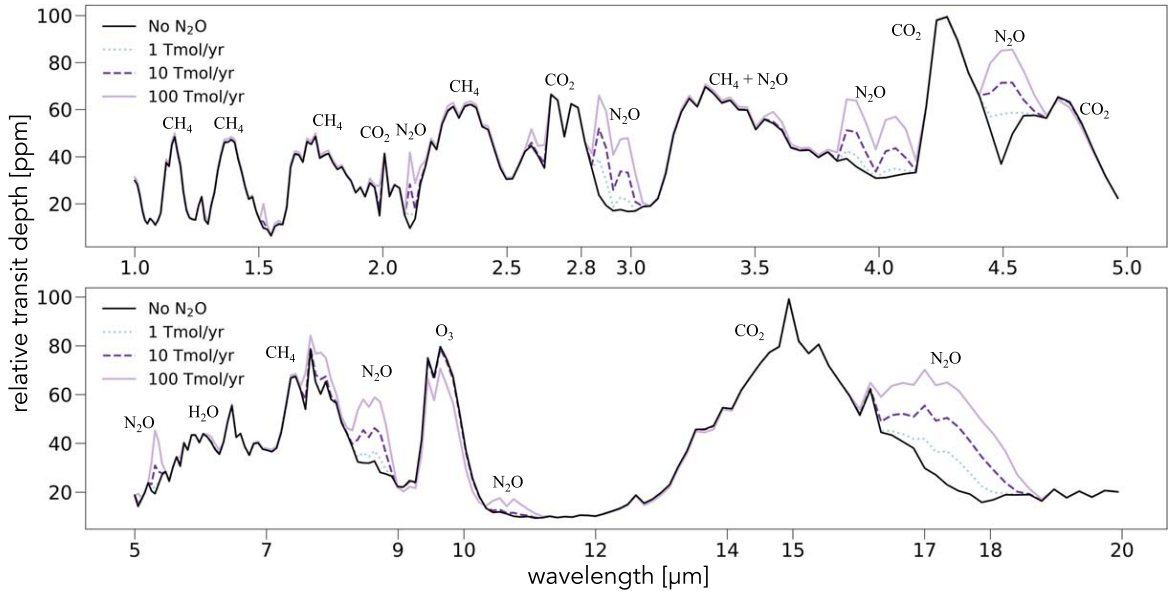

**Figure 11.** A simulated transmission spectrum of TRAPPIST-1e with N$_2$O surface fluxes of 1, 10, and 100 Tmol yr$^{-1}$, assuming the same gas profiles as shown in Figure 8(d), simulated with PSG. Major spectral features are noted. The top panel shows wavelengths accessible to the JWST NIRSpec instrument, while the bottom panel shows wavelengths accessible to the JWST MIRI instrument.

### 4.3. TRAPPIST-1e Transit Test Case with JWST

We used PSG to simulate the NIR and MIR transmission spectra of our Earthlike TRAPPIST-1e and to determine the number of transits necessary to detect N$_2$O with the JWST NIRSpec configuration. When calculating transmission spectra, we adopt the TRAPPIST-1e atmospheric profiles shown in Figure 9 and the planetary parameters in Agol et al. (2021).

Figure 11 shows the simulated transmission spectrum of the Earthlike TRAPPIST-1e with surface N$_2$O fluxes of 0, 1, 10, and 100 Tmol yr$^{-1}$ from 1 to 20 μm. The bottom panel of Figure 11 is directly comparable to the emission spectrum of TRAPPIST-1e from Figure 10, showing the complementarity of these two observing modes. The major bands in the NIR are located at 2.25, 2.9, 4.0, and 4.5 μm, with relatively weaker bands at 1.5, 1.6, 1.7, 1.8, 2.6, and 3.7 μm. The overlap with the CH$_4$ absorption strongly attenuates the impacts of many of these bands, so their impacts on the transmission spectra are not entirely the result of their intrinsic opacities, but also the overlap with the CH$_4$ bands centered at 1.4, 1.7, 2.3, and 3.4 μm. At a modest N$_2$O flux of 1 Tmol yr$^{-1}$, features are only apparent at 2.1, 2.9, 4.0, and 4.5 μm. At N$_2$O fluxes of 10–100 Tmol yr$^{-1}$, the relative transit depths of these bands become comparable to those of CH$_4$ and CO$_2$. The major bands in the MIR transmission spectrum are at 7.8, 8.5, and 17 μm. As in the emission spectrum, the 7.8 μm band is effectively hidden by the 7.7 μm CH$_4$ band. Also, as seen in the emission spectrum, the 9.65 μm O$_3$ band becomes truncated, due to more efficient photochemical destruction in the highest N$_2$O flux case.

**Table 2**
JWST Detectability Calculations for N$_2$O Flux Scenarios

| Flux Case | Feature (μm) | Transits for 3σ | Transits for 5σ |
|---|---|---|---|
| 1 Tmol yr$^{-1}$ | 2.10–2.16 | >1000 | >1000 |
|  | 2.86–2.97 | 148 | 409 |
|  | 3.89–4.09 | >1000 | >1000 |
|  | 4.47–4.52 | 212 | 587 |
| 10 Tmol yr$^{-1}$ | 2.10–2.16 | 140 | 388 |
|  | 2.86–2.97 | 35 | 97 |
|  | 3.89–4.09 | 109 | 303 |
|  | 4.47–4.52 | 81 | 223 |
| 100 Tmol yr$^{-1}$ | 2.10–2.16 | 36 | 98 |
|  | 2.86–2.97 | 13 | 36 |
|  | 3.89–4.09 | 25 | 68 |
|  | 4.47–4.52 | 41 | 114 |

We calculate the detectability of the NIR N$_2$O bands with JWST for all cases. We use the PSG JWST instrument simulation for the NIRSpec Prism to calculate the signal-to-noise ratio (S/N) for each major band, and subsequently calculate the number of transits needed to achieve an S/N ratio sufficient for detection to 3σ and 5σ. We add out-of-transit noise, using a transit to out-of-transit time ratio of 1:3, which lowers the S/N by a factor of ∼1.16. This ratio was selected to match the planned observations of TRAPPIST-1e with JWST (JWST GTO program 1331; Lewis et al. 2017). Our tabulated detectability estimates are given in Table 2 for the bands centered at 2.1, 2.9, 4.0, and 4.5 μm.

We find that for fluxes of 10–100 Tmol yr$^{-1}$, N$_2$O is plausibly detectable on TRAPPIST-1e (assuming that it is an Earthlike world). For example, considering a flux of 100 Tmol yr$^{-1}$, the 2.9 μm band of N$_2$O would be detectable to 3σ in 13 transits and 5σ in 36 transits. Given a flux of 10 Tmol yr$^{-1}$, N$_2$O could be detectable at 3σ in 35 transits. We find that N$_2$O is not likely to be detectable for fluxes around or less than 1 Tmol yr$^{-1}$, consistent with previous results (Wunderlich et al. 2019). Figure 12 shows our 100 Tmol yr$^{-1}$ TRAPPIST-1e scenario with realistic noise from PSG for 40 transits. The gray shaded area indicates the 3$\sigma$ noise envelope, while the bottom panel illustrates the selective impact of N$_2$O on the planetary spectrum.

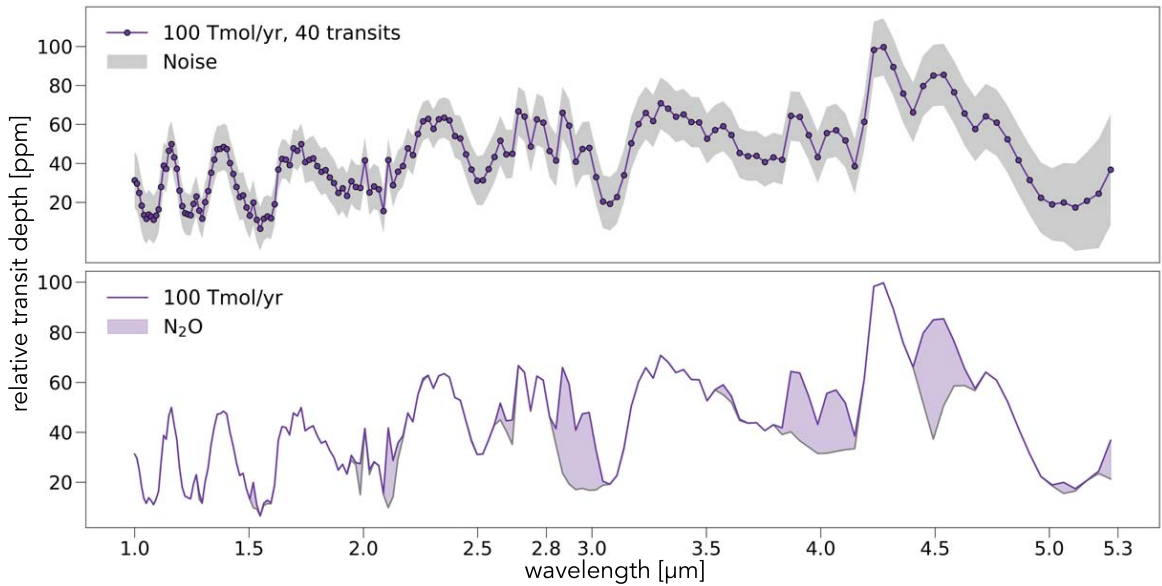

**Figure 12.** A simulated transmission spectrum of the Earthlike TRAPPIST-1e case, with a flux of 100 Tmol yr$^{-1}$ N$_2$O. The top panel shows simulated 5$\sigma$ error envelopes, as calculated with PSG, assuming 40 coadded transits with the JWST NIRSpec instrument. The bottom panel shows the influence of N$_2$O on the spectral features of the top panel.

We considered the detectability of the 7.8 and 8.5 $\mu$m bands with the MIRI-LRS instrument, but found that the number of transits to detect even the 100 Tmol yr$^{-1}$ case at 5$\sigma$ exceeded the nominal five-year mission lifetime of JWST (results not shown). We also considered the detectability of all bands with Origins MISC-T, but did not find a notable advantage over JWST NIRSpec in the NIR or JWST MIRI-LRS in the MIR (results not shown).

As was the case for the simulated MIR spectra presented in Section 4.2, the primary difference between the N$_2$O transmission signatures that we predict here and the estimates from other researchers simulating synthetic transmission spectra of Earthlike planets with N$_2$O (e.g., Wunderlich et al. 2019, 2021) is due to our consideration of larger potential biogenic fluxes. A globally averaged Earth flux applied to TRAPPIST-1e would result in the N$_2$O spectral features falling below the 1 Tmol yr$^{-1}$ spectrum in Figure 11.

## 5. Discussion

### 5.1. Possible N$_2$O Flux Regimes

Our results demonstrate that the detection of N$_2$O on O$_2$-rich terrestrial exoplanets is plausible in the near-to-intermediate future for scenarios where the N$_2$O flux approaches 5%–10% of the modern Earth's global denitrification flux or greater. How likely is this to happen? It is difficult to extrapolate from one data point, but both the energetics of nitrogen metabolism and the biogeochemical evolution of Earth's oceans and atmosphere both suggest that such a scenario is very possible.

In the extreme case, life on an exoplanet may not have evolved the nitrous oxide reductase enzyme that facilitates the last step of the denitrification process (N$_2$O $\to$ N$_2$). This step, while thermodynamically favorable, has the highest kinetic barrier (Pauleta et al. 2013; Carreira et al. 2017), so it is less evolutionary advantageous to evolve this step instead of any other. We have demonstrated that the atmosphere of a planet with even a maximally productive biosphere would not become dominated by N$_2$O, due to the efficiency of N$_2$O photolysis. In effect, the last step of the denitrification cycle (N$_2$O $\to$ N$_2$) on such a planet would be accomplished abiotically, via photolysis, and the atmosphere would never be placed into a chemical runaway (Ranjan et al. 2022). However, we have found that the N$_2$O concentrations of such a world would be high compared to those of present-day Earth—between tens and thousands of ppm N$_2$O, depending on the host star.

An N$_2$O concentration this high would constitute a large unexploited chemical disequilibrium. An unexploited disequilibrium can be interpreted as a biosignature or antibiosignature, depending on the context (Wogan & Catling 2020). In this case, N$_2$O concentrations of tens or even thousands of ppm are not conceptually different from the unexploited equilibria on Earth today and in the geologic past—the result of a highly productive photosynthetic biosphere. For example, CH$_4$ is out of equilibrium with O$_2$ on Earth, which has long been interpreted as a biosignature (Lovelock 1965, 1975; Hitchcock & Lovelock 1967). (It is not kinetically efficient for life to pull large amounts of CH$_4$ out of the atmosphere to react with O$_2$, even though it is thermodynamically favorable.) Methane on the Archean Earth was out of equilibrium with the CO$_2$ in the atmosphere (Krissansen-Totton et al. 2018), and the CH$_4$ concentrations on Archean Earth may have been in the range of hundreds to thousands of ppm or greater (Arney et al. 2016; Olson et al. 2018a; Robinson & Reinhard 2018), which is comparable to the maximum N$_2$O concentrations that we find here.

Even if the nitrous oxide reductase enzyme (or its analog) had evolved on an exoplanet, N$_2$O fluxes may still be high, depending on the environmental context. Trace nutrient limitation may impact the viability of biological N$_2$O reduction to N$_2$, which likely depends on enzymes (as it does on Earth). As an example, the nitrous oxide reductase enzyme requires copper catalysts (Carreira et al. 2017), and if copper were sharply limited, a much greater proportion of the total denitrification flux would be released into the atmosphere as $N_2O$ versus $N_2$. The initial copper inventory may be limited on an exoplanet compared to Earth, based on the balance of pebble and planetary accretion to larger impacts (Mahan et al. 2018) or the scatter in stellar metallicities (Delgado Mena et al. 2017). Alternatively, widespread ocean euxinia (simultaneous anoxia and elevated levels of $H_2S$ in the water column) could dramatically lower the availability of copper as a micronutrient, which has been proposed for the earlier periods of Earth's evolutionary history (Buick 2007). Euxinic oceans are more likely to occur at lower oxygen levels, which work against $N_2O$ lifetimes and, by extension, predicted abundances. However, euxinic environments are not completely unknown even on Earth today (Lyons & Severmann 2006), and they could be more extensive on a well-oxygenated exoplanet, given different balances between volcanic sulfur ($H_2S$) fluxes, ocean volumes, and the number and nature of anoxic basins. Moreover, intermediate oxygenation states of 20%–50% PAL would both favor euxinia and maximize denitrification fluxes, while conferring only a modest impact on $N_2O$ lifetimes, due to reduced $O_2$ shielding. The photolysis dependence of $N_2O$ on $O_2$ levels is substantially reduced for late K- and M-dwarf hosts versus Sun-like F, G, and early K hosts, so lower $O_2$ levels will be less impactful on $N_2O$ abundances for the planets orbiting these stars.

Other planetary environmental conditions may impact $N_2O$ production, including pH effects (Chen et al. 2015). The ocean pH of planets in the habitable zone may vary over a wide range, encompassing acidic (pH < 7) as well as basic (pH > 7) ocean chemistries, based on the required $CO_2$ to maintain clement conditions and equilibrate the carbonate-silicate thermostat (Schwieterman et al. 2019b; Krissansen-Totton & Catling 2020). Planetary environmental conditions may impact the community composition of the organisms that are involved in the nitrogen cycle, which will affect the production and loss of biological or abiotic products that can have an inhibitory impact on nitrous oxide reductase, such as CO, $CN^-$, $I^-$, $C_2H_2$, and $N_3^-$ (Kristjansson & Hollocher 1980; Koutný & Kučera 1999; Paraskevopoulos et al. 2006). An important task moving forward will be to more definitively evaluate the conditions of ocean biogeochemistry that would be most likely to lead to elevated $N_2O$ fluxes, and to establish how widespread these might be for habitable worlds.

We briefly note that agricultural activity has resulted in enhanced production of fixed nitrogen on Earth, which has in turn increased the rate of terrestrial $N_2O$ production (Tian et al. 2020). Haqq-Misra et al. (2022) proposed that anomalously high $NH_3$, $N_2O$, and $CH_4$ in combination may serve as a "technosignature" for extensive planetary agriculture, and applied this concept to various scenarios for future population growth on Earth. Our results suggest a substantial overlap between the predicted $N_2O$ flux and mixing ratios for robust agriculture and an ocean biosphere where a large fraction of the denitrification flux is released as $N_2O$. However, we do not predict that this would lead to a simultaneous evolution of $NH_3$, which underpins the arguments in Haqq-Misra et al. (2022). Moreover, our predicted photochemical flux–abundance relationships for FGKM stars can be applied to modeled agricultural production of $N_2O$, and could therefore inform future studies similar to that of Haqq-Misra et al. (2022).

Ultimately, a robust production flux of $N_2O$ is essential for both the practical considerations of detectability and our ability to infer an explicit link to biological production, since abiotic processes can produce $N_2O$ at low levels (e.g., Schumann & Huntrieser 2007; also see the text below). To rule out processes that may generate false positives, all biosignature gases, including $N_2O$, must be evaluated in terms of the planetary context (Meadows et al. 2018b; Krissansen-Totton et al. 2022). Figure 13 is a concept graphic that illustrates planetary scenarios in which $N_2O$ can either be interpreted as a biosignature or ruled out as a clear biosignature, given the planetary context or other complementary spectroscopic observables. We discuss these scenarios in detail in the subsections below.

### 5.2. Stellar Activity and False Positives

Stellar activity and solar proton events (SPEs) have long been known to produce NO in Earth's atmosphere, via the secondary production of energetic electrons (Crutzen et al. 1975). These NO precursors could be a potential source of abiotic $N_2O$ in some planetary atmospheres with sufficient H-bearing species, via the reaction $NO + NH \rightarrow N_2O + H$ (Airapetian et al. 2016, 2020). Airapetian et al. (2016) found that $N_2O$ could have been generated at the ppb level on the anoxic and weakly reducing Hadean Earth by a younger, more active Sun. An active Sun would produce substantially higher fluxes of energetic protons and electrons, which can split $N_2$ into N in the upper atmosphere (normally the photolysis of $N_2$ is limited by the paucity of solar photons energetic enough to break the $N \equiv N$ triple bond). Subsequent photochemical reactions can produce $N_2O$, e.g., $N(^4S) + NO_2 \rightarrow N_2O + O$, in addition to $NO + NH \rightarrow N_2O + H$. More recent refinements to this model have predicted ground-level $N_2O$ concentrations as high as $\sim$1 ppm and stratospheric concentrations as high as $\sim$1000 ppm, which could have had notable impacts on the Hadean or early Archean climate (Airapetian et al. 2018, 2020). The highest predicted (column-integrated) abundances are comparable to those of our 1–10 Tmol yr$^{-1}$ models.

How could we distinguish between biological $N_2O$ and $N_2O$ that is generated from stellar SPEs? The characterization of the host star is essential. A critical consideration is whether the star is young and magnetically active, presenting qualities associated with SPEs and coronal mass ejections (CMEs), which may be indicated by EUV and X-ray and extreme-ultraviolet (XUV) flares (Hu et al. 2022). A star with similar activity to the modern Sun will not produce meaningful abiotic atmospheric $N_2O$. Older, less magnetically active stars are also more enticing targets from the perspective of life detection, due to the additional time allowed to evolve a complex biosphere (Turnbull & Tarter 2003). The characterization of the host star required for this activity is likely to be less onerous than the detection of biosignature gases in orbiting terrestrial planets. Forward ion photochemical modeling, such as that performed by Airapetian et al. (2016), can make predictions of the abiotic $N_2O$ generation potential, given observed time-dependent stellar inputs. Finally, age and activity estimates for nearby Sun-like stars with directly imageable habitable zones are known (Turnbull & Tarter 2003; Reid et al. 2007; Turnbull 2015), with most being >$\sim$3 Gyr or older, and therefore less likely to be active and plausible candidates for efficient abiotic production of $N_2O$ on orbiting terrestrial planets.

Spectral discriminants, such as HCN, could also be used to fingerprint the abiotic generation of $N_2O$ in weakly reducing anoxic atmospheres that contain $CH_4$. The photochemical modeling of abiotic $N_2O$ generation predicts large quantities of other spectrally active gases, including HCN (Airapetian et al. 2016, 2020). The production of abiotic HCN and $N_2O$ are strongly linked, because the energetic particles that generate the NH radicals that lead to $N_2O$ production also generate the $CH_x$ radicals that are necessary for HCN production (Airapetian et al. 2016). HCN absorbs most notably at 1.5, 2.5, 3.0, 7.0, and 14 $\mu$m, with weaker bands at 1.4, 1.85, and 4.8 $\mu$m. The role of HCN in this case would be similar to that proposed for CO and $O_4$, which can fingerprint abiotic $O_2$ (Harman et al. 2015; Schwieterman et al. 2016; Meadows et al. 2018b). Similarly, CMEs generate large amounts of potentially detectable $HNO_3$ at 5.9, 7.5, and 11.3 $\mu$m for $O_2$-rich atmospheres (Tabataba-Vakili et al. 2016). The third panel of Figure 13 provides a conceptual illustration of how we may distinguish the abiotic $N_2O$ due to SPEs from true positive $N_2O$ biosignatures.

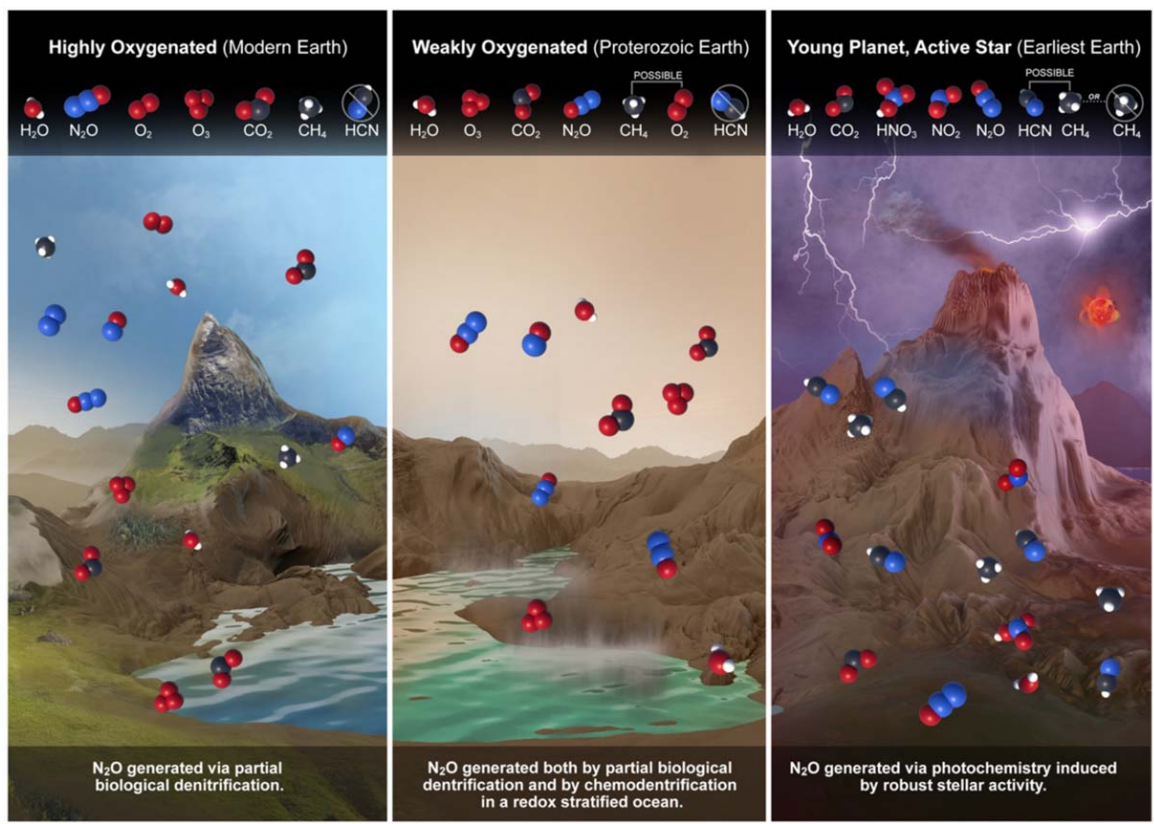

**Figure 13.** A concept image illustrating the interpretability of $N_2O$ as a biosignature in the context of the planetary environment. The left panel illustrates a scenario like the modern Earth, with a high-oxygen atmosphere and $N_2O$ generated overwhelmingly via partial biological denitrification. In this case, the simultaneous presence of $O_2$, $O_3$, $N_2O$, and $CH_4$ indicates a strong chemical disequilibrium. The middle panel illustrates a weakly oxygenated planetary environment like the Proterozoic Earth, with $N_2O$ generated both by partial biological denitrification and by chemodenitrification of nitrogenous intermediates (likely substantially biogenic) in a redox stratified ocean. In this case, molecular oxygen ($O_2$) and methane may have concentrations that are too low to detect directly, but detectable $N_2O$ and $O_3$ would be a strong biosignature. A false positive is unlikely, because an abiotic $O_2$ atmosphere would be unstable in combination with a reducing ocean. The right panel illustrates the most likely false-positive scenario, where an active star splits $N_2$ via SPEs, resulting in photochemically produced $N_2O$. This scenario would predict additional photochemical products, such as HCN, that would be indicative of abiotic origins. Stellar characterization would confirm the magnitude of the stellar activity. Vigorous atmospheric production of $NO_x$ species could be inferred from spectrally active $NO_2$.

We note that predictions of HCN in conjunction with abiotic $N_2O$ are predicated on the presence of the HCN precursor $CH_4$. According to Airapetian et al. (2016), this $CH_4$ would directly shield against $N_2O$ photolysis, which would otherwise be a strong sink for $N_2O$ (see Section 3.2). However, while Airapetian et al. (2016) argued that $CH_4$ provides shielding to wavelengths <230 nm, other sources for $CH_4$ cross sections indicate a threshold cutoff for strong $CH_4$ absorption at $\sim$ > 150 nm (Keller-Rudek et al. 2013). Future work should investigate the impact of differing implementations of $CH_4$ cross sections on abiotic $N_2O$ production.

The production of $N_2O$ by SPEs generates a potentially identifiable altitude profile of $N_2O$. Since $N_2O$ is preferentially generated in the stratosphere, concentrations are substantially higher than in the troposphere by a factor of $\sim$1000 (Airapetian et al. 2020). This prediction is a notable deviation from the appearance of $N_2O$ profiles given a surface source (e.g., Figure 9) where concentrations are maximal in the troposphere. If altitude-dependent $N_2O$ (or HCN) profiles can be retrieved, a surface biological source of $N_2O$ could be distinguished from an atmospheric one.

### 5.3. Is Chemodenitrification a Possible False Positive?

Chemodenitrification can produce abiotic $N_2O$ via reduction of $NO_x$ species by ferrous iron (e.g., $4Fe^{2+} + 2NO_2^- + 5H_2O \rightarrow 4FeOOH + N_2O + 6H^+$; Jones et al. 2015). Chemodenitrification is observed in one of the very rare metabolically inert environments on Earth, the hypersaline Don Juan Pond in Antarctica (Samarkin et al. 2010; Peters et al. 2014). Chemodenitrification may have played a role in maintaining clement climates on early Earth, when intermediate oxygenation levels may have allowed ferruginous ($Fe^{2+}$-rich) oceans and oxidized nitrogenous intermediates from denitrification and nitrification (NO, $NO_2^-$) to exist together in meaningful concentrations (Stanton et al. 2018). In this scenario, nitrogenous intermediates are biologically generated in an oxic upper ocean layer and subsequently mixed into a deeper $Fe^{2+}$-rich anoxic layer, where chemodenitrification can take place. Stanton et al. (2018) find that in the end-member chemodenitrification scenario, where all nitrogenous intermediates are released as $N_2O$ via the abiotic oxidation of ferrous iron, parts per million $N_2O$ can be achieved, which is comparable to our maximum $N_2O$ concentrations for the Sun, assuming modern denitrification fluxes. A critical question is whether this scenario is a potential false positive for $N_2O$ biosignatures. The short answer is no, because it relies on the existence of a large pool of nitrogen oxides in the ocean, which is out of equilibrium with a reducing ocean. The pool of nitrogenous intermediates is a product of biological nitrification and denitrification processes. In effect, chemodenitrification in this context short-circuits biological denitrification for the step between $NO_2^-$ and NO or the step between NO and $N_2O$, with a more efficient abiotic step that produces abiotic $N_2O$ from biotic N substrates. The large flux of $N_2O$ would thus fingerprint (de)nitrification fluxes that are sufficiently large to be attributed to a biosphere, albeit indirectly, and so should not complicate biosignature interpretations.

Another important consideration is whether large pools of abiotic aqueous nitrogen oxides coexist with a ferruginous ocean. We consider this unlikely. The biological nitrification of $NH_4^+$ to $NO_3^-$ ($NH_4^+$ $NH_2OH \rightarrow NO \rightarrow NO_2^-$ $NO_3^-$) requires oxygen. Other sources of $NO_x$ species, such as lightning, are likely to be much weaker, as they have been on Earth since the Archean (Navarro-Gonzalez & Mvondo 2001). In an abiotic $O_2$ atmosphere, perhaps $NO_x$ species produced by lightning could accumulate over time to saturation in the ocean. However, in this scenario, there would likely be no $Fe^{2+}$ intermediate to reduce these $NO_x$ species to $N_2O$. Chemodenitrification requires both oxic and anoxic regions within the ocean, which requires substantial fluxes of both oxidants (e.g., $O_2$) and reductants (e.g., $Fe^{2+}$). An atmosphere that has accumulated abiotic $O_2$ would have done so because abiotic $O_2$ production has completely overwhelmed reductant sinks. Conversely, a flux of reductants sufficiently large to maintain an anoxic ocean layer would be inimical to the maintenance of an abiotic $O_2$ atmosphere, which would disappear over relatively short geologic timescales. Indeed, past predictions of abiotic $O_2$ accumulation assumed an ocean saturated in $O_2$, as otherwise this condition would not occur (Hu et al. 2012, 2020; Domagal-Goldman et al. 2014; Tian et al. 2014; Harman et al. 2018). The presence of $N_2O$, which lies at an intermediate redox state to $O_2$ and $Fe^{2+}$, is an indirect indicator of ocean disequilibrium that excludes abiotic $O_2$ atmospheres that are stable over geologic time.

Recent laboratory experiments of haze formation under Proterozoic-like $N_2$–$CO_2$–$CH_4$–$O_2$ atmospheric conditions have suggested the possibility of nitrogen fixation facilitated by the presence of haze particles (Hörst et al. 2018). If this process were very efficient in a Proterozoic-like atmosphere, $NO_x$ species could be delivered to the ocean, which would be subjected to denitrification via biological metabolism or chemodenitrification in an $Fe^{2+}$-rich ocean, with the possible evolution of $N_2O$. However, the very presence of the haze requires the simultaneous presence of $O_2$, $CH_4$, and $N_2$, a strong atmospheric and oceanic disequilibrium that correspondingly requires large fluxes of $O_2$ and $CH_4$ (Zerkle et al. 2012; Krissansen-Totton et al. 2016, 2018). In this case, the resulting $N_2O$ remains an indirect indicator of this ocean and atmospheric disequilibrium, which is ultimately the result of biology. Furthermore, this $N_2O$ is likely a more observable biosignature than the trace levels of $O_2$ (2 ppm to 0.2%) that are required to trigger nitrogen fixation in the haze (Hörst et al. 2018).

Another possibility is a large lightning flux on a planet with a slightly reducing atmosphere and ocean and high $CO_2$ concentrations ($>\sim100$ PAL), such as those hypothesized for the early Archean or late Hadean Earth, along with an intermittently warmer and wetter climate and an increased convection rate (Catling & Zahnle 2020). Wong et al. (2017) proposed that lightning on the early Earth could have generated relevant levels of $NO_x$ species, which could accumulate to prebiotically relevant concentrations in the ocean. However, Ranjan et al. (2019) argued that these $NO_x$ species would be efficiently converted to $N_2$ or $N_2O$ gas, partially via reaction with $Fe^{2+}$ in the ocean. We also consider it unlikely that these lightning-induced $NO_x$ species would ultimately lead to detectable $N_2O$ in a prebiotic Earthlike atmosphere, because such an atmosphere would lack the $O_2$ (and $O_3$) shielding that facilitates large $N_2O$ buildup. The $NO_x$ production rates predicted by Wong et al. (2017) and Ranjan et al. (2019) depend on atmospheric assumptions, such as $pCO_2$ and $pN_2$. The largest predicted lightning-induced $NO_x$ flux from Wong et al. (2017) for $pCO_2 = 1$ bar is $6.5 \times 10^8$ molecules $cm^{-2}$ $s^{-1}$ ($\sim 0.2$ Tmol $yr^{-1}$). If we assume that 100% of this flux is converted into $N_2O$, we predict a $\sim$1–100 ppb level of $N_2O$ for all stars, even at an atmospheric $pO_2$ of 1% PAL (Figures 4 and 5). The calculated $N_2O$ is likely to be even lower with no $O_2$ and its attendant shielding effects. (Note that we do not consider shielding from 1 bar of $CO_2$.) $NO_2$ is a direct product of lightning and is spectrally active at 6.2 $\mu$m (Gordon et al. 2022). Future work should quantify the limits of $N_2O$ detectability for lifeless atmospheres and spectral discriminants for false positives in more detail, but at present we consider potential false positives for $N_2O$ as a biosignature to be either implausible or readily diagnosable when viewed in context (Figure 13).

### 5.4. Potential for Seasonal $N_2O$ Signatures

Seasonal variations in atmospheric gases, including $CH_4$, $CO_2$, or $O_2$/$O_3$, could potentially be interpreted as a biosignature with the right planetary context (Meadows 2006, 2008; Olson et al. 2018b; Schwieterman 2018). Seasonality can be driven by planetary obliquity, as on Earth today, or by large orbital eccentricity (Gaidos & Williams 2004; Olson et al. 2018b). We do not explicitly explore $N_2O$ seasonality as a potential biosignature here, but our results are relevant for future work on this topic. Seasonal variations in atmospheric biosignature gases will be the result of the interplay between time-dependent production, time-dependent photochemical destruction, and temperature-dependent solubility (Olson et al. 2018b). The photochemical lifetime of $N_2O$ on Earth is around 120 yr (Prather et al. 2015), which is substantially greater than the lifetime of $CH_4$, which is about 10 yr (Voulgarakis et al. 2013). Correspondingly, the observed relative seasonal variability of $N_2O$ on modern Earth is markedly lower than that of $CH_4$, about $\sim$0.1%–0.3% for $N_2O$ (Jiang et al. 2007) compared to $\sim$1%–2% for $CH_4$ (Rasmussen & Khalil 1981; Dlugokencky et al. 2017), depending on latitude. The seasonal destruction of $N_2O$ is less impacted by temperature than $CH_4$, because $CH_4$ is primarily destroyed by OH radicals that are ultimately liberated from tropospheric $H_2O$ (Khalil & Rasmussen 1983). Warmer temperatures lead to more tropospheric $H_2O$, and thus more OH and less $CH_4$. In contrast, $N_2O$ is primarily destroyed by photolysis, with a smaller impact from $O(^1D)$ radicals that are sourced from $O_3$ rather than $H_2O$.

The greatest seasonal impact will occur when high seasonally variable biological fluxes are matched with high (potentially seasonally variable) photochemical destruction rates (low photochemical lifetimes). Eccentric planets are likely the best targets for seasonal variability, because seasonality driven by eccentricity can avoid many of the degeneracies that are expected on planets with seasonality driven by obliquity. In the case of obliquity, the signal of seasonality can be muted by the mixing of offsetting hemispheric trends, depending on viewing geometry (Olson et al. 2018b). Moreover, the decay and rise of atmospheric gases as a function of the changes in orbital position driven by eccentricity can be more directly linked to inferred global molecular fluxes, which is fundamental to biosignature interpretations.

Based on our results, the highest photochemical destruction rates for $N_2O$ will occur for low-$O_2$ planets orbiting F- or G-type stars (Figures 4 and 5). Indeed, biologically modulated seasonality in $O_2$ (and $O_3$) themselves, as explored by Olson et al. (2018b), would strongly contribute to this relationship. Of course, there is an interplay between the baseline concentration of a gas and its relative variability when considering detectability. We suggest that future studies of $N_2O$ seasonality should explore the phasing and interplay between seasonally dependent denitrification fluxes and seasonal $O_2/O_3$ concentrations for planets with intermediate oxygenation states orbiting F and G stars, as this set of criteria will maximize the chance of predicting observable $N_2O$ seasonality.

*5.5. Study Limitations and Alternative Atmospheric Scenarios*

In this study, we have examined $N_2O$ flux–abundance relationships and related photochemistry for Earth-size terrestrial planets with $N_2$–$O_2$ dominated atmospheres. We have tested the sensitivity of these flux–abundance relationships to changes in $O_2$ concentrations, from 1% to 100% PAL for FGKM stellar hosts (Figures 4–9; also see Figure A1). These conditions are motivated in part by the observation that robust biological $N_2O$ production on Earth requires the presence of oxygen to generate the necessary nitrogenous intermediates (e.g., see Figures 1–2), and $N_2$ is required as a reservoir of N to support the nitrogen cycle. However, we can consider alternative atmospheric scenarios.

The biological production of $N_2O$ on moderately reducing, Archean-like worlds might be facilitated by the presence of oxygen oases that are inhabited by oxygenic photosynthesizers. Oxygen oases have been proposed as possible features of the late Archean Eon on Earth (Anbar et al. 2007; Olson et al. 2013; Reinhard et al. 2013; Riding et al. 2014). However, the flux–abundance relationships of such worlds would fall below those predicted for our 1% PAL $O_2$ scenarios, illustrated in Section 3.2, due to the lack of shielding effects by $O_2$ and $O_3$.

We did not explicitly consider a moist greenhouse state, which may enhance convection and, potentially, lightning rates and $NO_x$ production (Tost et al. 2007; Sergeev et al. 2020). The production of $NO_x$ species, however, will not result in appreciable $N_2O$ buildup via chemodenitrification without the presence of a stratified ocean with both oxygen-rich and anoxic layers, such as existed during the Proterozoic Eon on Earth. As discussed in Section 5.3, a redox stratified ocean is unstable without continuous sources of oxidants and reductants, and $N_2O$ is an intermediate redox product of this likely biologically supported disequilibrium. Nonetheless, future work could explore the rate of $NO_x$ production on worlds with moist greenhouses, including the incorporation of differing circulation regimes on tidally locked planets (Sergeev et al. 2020). An end-member state would be a so-called "steam atmosphere" (e.g., Mousis et al. 2020). A steam atmosphere may be a poor candidate for searching for $N_2O$ biosignatures, however, because the deep atmosphere will contain a supercritical water layer above the surface—with temperature conditions that exclude life as we know it.

Conceivably, some level of $N_2O$ could be generated as a tertiary metabolic product on anoxic planets, potentially including worlds with $H_2$-dominated atmospheres, as suggested by Seager et al. (2013a, 2013b). The generation of $N_2O$ on a highly reducing world would require the presence of terminal electron acceptors (oxidants), which would likely be scarce in such an environment. However, if the $N_2O$ is nonetheless generated in sufficient abundance, it would be more easily detectable in transit transmission observations, due to the enhanced scale height (Seager et al. 2013a, 2013b). Future work would be needed to fully explore the flux–abundance relationships and, as a consequence, detectability estimates for $N_2O$ in $H_2$-dominated planetary atmospheres.

Our study used a 1D photochemical model and assumed vertical mixing consistent with an Earth average. Exoplanets, particularly those with different circulation regimes, could have substantially different vertical mixing efficiencies (Zhang & Showman 2018). Enhanced vertical mixing would lead to larger $N_2O$ photolysis rates and less $N_2O$ for any given surface flux. We quantify the impact of adjusting the eddy diffusion profile upward by a factor of 10 on a subset of our flux–abundance calculations in Appendix Figure A4, finding small to modest declines in the predicted surface $N_2O$ mixing ratios. 3D models can more accurately model atmospheric circulation, particularly for tidally locked planets, which may result in hemispheric gradients (Chen et al. 2018, 2019; Yates et al. 2020). Alternative predictions for ozone and other trace gas concentrations would also impact chemistry-climate feedbacks (Gómez-Leal et al. 2019; Yates et al. 2020). We note that Chen et al. (2018) found that the day–night hemisphere contrasts in mixing ratios on synchronously rotating planets differ by only ~20% for standard biosignatures gases like $CH_4$ and $N_2O$, so the overall impact of the simplifications inherent in 1D modeling may be small compared to other factors (such as stellar type and bulk atmospheric composition). Nonetheless, future work should more fully explore the impacts of varied circulation regimes on $N_2O$ flux–abundance relationships and potential detectability.

Finally, we further emphasize some of the fundamental assumptions that have been made in this study that will not apply in all scenarios. We have assumed that all modeled planets had $N_2$ atmospheres with a nitrogen reservoir that would not be limiting for biological nitrification and denitrification. $N_2$ is an expected volcanic product on terrestrial planets (Schaefer & Fegley 2010), and stellar nitrogen abundances closely track carbon abundances and overall metallicity in the solar neighborhood (Hinkel et al. 2014). However, the primordial origin of nitrogen during the formation of the planets in the inner solar system, particularly the relative importance of delivery from volatile rich bodies formed outside the ice line, is the subject of ongoing debate (Marty 2012; Alexander et al. 2017; Grewal et al. 2021), and therefore the likelihood and efficiency of this delivery mechanism on exoplanets is unresolved. Planets with substantially lower nitrogen abundances by several orders of magnitude compared to Earth could have biospheres that are limited by the availability of nitrogen. In this case, the capacity of a planetary biosphere to produce any biosignature, including $N_2O$, would be highly suppressed. On the other hand, $N_2$-rich planets may have substantially larger atmospheric masses and $N_2$ partial pressures compared to Earth. We quantify the impact of $N_2$ partial pressure between 0.5 and 10 bars in Appendix Figure A3, finding negligible declines in the predicted $N_2O$ for the 0.5 bar scenario and modest increases in the predicted $N_2O$ for the 10 bar cases.

Habitable planets in the outer habitable zones of their host stars may be expected to have vastly larger $CO_2$ abundances, due to the coupling between greenhouse physics and the carbonate-silicate weathering thermostat, possibly as high as 5–20 bars (Kasting et al. 1993; Kopparapu et al. 2013; Rushby et al. 2018; Schwieterman et al. 2019b). We did not model high $CO_2$ atmospheres, in part because we considered Earthlike planets with Earthlike instellations, which would be inconsistent with these very high $CO_2$ abundances. We predict that lower instellations would result in less $N_2O$ photolysis, and therefore higher $N_2O$ concentrations, at a given flux. Additionally, $CO_2$ would provide additional shielding effects at wavelengths shorter than ~200 nm, which could enhance $N_2O$ concentrations, assuming constant $O_2$ and $O_3$ shielding. However, the likelihood of oceanic and atmospheric oxygenation on outer habitable-zone planets, and therefore the potential for robust $N_2O$ production and photochemical stability, has yet to be explored. We reserve the studying of these high-$CO_2$ scenarios for future work.

We modeled Earthlike atmospheres on planets orbiting M-dwarf stars, though the capacity of M-dwarf planets in the continuous habitable zone to retain atmospheres in the face of a long-lived superluminous pre-main-sequence phase and attendant XUV radiation is unresolved (Ramirez & Kaltenegger 2014; Luger & Barnes 2015; Tian 2015). Mass–radius studies of the TRAPPIST-1 system hint that some of these planets retain atmospheres (Grimm et al. 2018; Agol et al. 2021); however, this question will only be definitively resolved by near-future observations by JWST or future telescopes. If M-dwarf planets tend not to retain atmospheres, our results are still valid for the FGK stellar hosts modeled.

### 5.6. Future Prospects for Detecting $N_2O$ on Exoplanets

The detectability of $N_2O$ spectroscopic biosignatures will also depend on the observing mode (e.g., transit transmission or direct-imaging spectroscopy), the accessible wavelength range, the characteristics of the observed system, and the specifics of the instrument(s) and observing strategy (e.g., Fujii et al. 2018). Because the strongest bands of $N_2O$ lie in the NIR (e.g., 2.9 and 4.5 $\mu$m) and MIR (e.g., 7.8 and 8.5 $\mu$m), these are more amenable targets for observing techniques that rely on IR transit transmission and/or emission spectroscopy (Line et al. 2019). Reflected light direct-imaging missions, such as the 6 m IR/O/UV telescope recommended by Astro2020 (National Academies of Sciences, Engineering, and Medicine 2021), will be unlikely to detect $N_2O$, due to the weak $N_2O$ bands shortward of 2 $\mu$m near the expected cutoff for such a mission (The LUVOIR Team 2019; Gaudi et al. 2020). JWST is unlikely to detect $N_2O$ on transiting planets with abundances or fluxes like those of Earth today (e.g., Pidhorodetska et al. 2020; Wunderlich et al. 2021). However, we have shown that $N_2O$ is plausibly detectable on TRAPPIST-1e, assuming an $N_2$–$O_2$ atmosphere and a biosphere that emits a large fraction of its total denitrification flux as $N_2O$. Moreover, intermediate-to-high levels of $N_2O$ could be detected in emitted light by 30 m ground-based telescopes (López-Morales et al. 2019) or by a space-based IR interferometer, such as the LIFE concept mission (Quanz et al. 2022; Alei et al. 2022). Future cooled IR missions may plausibly target the $O_3 + N_2O$ biosignature couple, as investigated by the Origins concept mission report (Meixner et al. 2019). Snellen et al. (2017) propose using a novel high-pass spectral filtering technique to detect the 15 $\mu$m $CO_2$ band on Proxima Centauri b using JWST's MIRI in Medium Resolution Spectrograph mode. Planet/star contrast ratios are particularly favorable at MIR wavelengths. Conceivably, if this technique can successfully reveal atmospheric $CO_2$, a similar technique could be used to target the 17 $\mu$m $N_2O$ band. We suggest that future work quantifying the detectability of biosignature gases should consider the possibility of the elevated $N_2O$ levels that we explore here.

### 6. Conclusions

We have conducted a systematic study of $N_2O$ flux–abundance relationships for $N_2$-dominated atmospheres over a large range of surface molecular fluxes (0.01–100 Tmol yr$^{-1}$), for $O_2$ abundances ranging from weakly to fully oxygenated (1%–100% PAL), and for potential host stars spanning the entire main sequence, from F4V to M8V. We used the biogeochemical model cGENIE to inform the maximum plausible $N_2O$ fluxes for an Earthlike biosphere, which could be 1–2 orders of magnitude larger than those on present-day Earth, assuming nutrient-rich oceans and evolutionary or environmental conditions that limit the last step in the denitrification process. Even for maximal biospheric $N_2O$ fluxes of 100 Tmol yr$^{-1}$, an Earthlike atmosphere will never enter an $N_2O$ runaway, but would attain much larger concentrations than those found on Earth today. We find that late K-dwarf and inactive early M-dwarf stars can maintain the highest $N_2O$ levels at any given surface flux, potentially exceeding 1000 ppm. We show that for $N_2O$ fluxes of 10–100 Tmol yr$^{-1}$, JWST could detect $N_2O$ at 2.9 $\mu$m on TRAPPIST-1e within its mission lifetime. In thermal emission spectra, $N_2O$ features at 7.8 and 8.5 $\mu$m become comparable to other gaseous features, such as $O_3$ at ~1 Tmol yr$^{-1}$ for late K dwarfs and 1–10 Tmol yr$^{-1}$ for G dwarfs. Terrestrial planets orbiting K-dwarf stars are particularly appealing targets for $N_2O$ searches with future MIR missions, due to favorable planet–star angular separations and because $N_2O$ fluxes of only 2 to 3 times those of Earth's modern global average can produce $N_2O$ signatures comparable to those of $O_3$. Only the 8.5 $\mu$m band of $N_2O$ is likely to be accessible in thermal emission for M dwarfs, due to the overlap from an enhanced 7.7 $\mu$m $CH_4$ band, but the 8.5 $\mu$m feature becomes comparable to $O_3$ for $N_2O$ fluxes between 1 and 10 Tmol yr$^{-1}$. Fully abiotic sources of $N_2O$ are limited, and false positives can be identified by relevant astrophysical and planetary context, including complementary spectral signatures.

The authors gratefully acknowledge financial support from the NASA Exobiology program through grant Nos. 80NSSC20K1437 and 80NSSC19K0461. The authors also gratefully acknowledge support from the NASA Interdisciplinary Consortia for Astrobiology Research (ICAR) Program via the Alternative Earths team, with funding issued under grant No. 80NSSC21K0594, and from the Consortium on Habitability and Atmospheres of M-dwarf Planets (CHAMPS) team, with funding issued under grant No. 80NSSC21K0905. This work has made use of tools developed by the Virtual Planetary Laboratory, which is a member of the NASA Nexus for Exoplanet System Science and funded via NASA Astrobiology Program Grant No. 80NSSC18K0829. T.J.F., S.T.B., and J.S. C. acknowledge support from the GSFC Sellers Exoplanet Environments Collaboration (SEEC), which is funded in part by the NASA Planetary Science Divisions Internal Scientist Funding Model. S.T.B. and J.S.C are also supported by NASA under award No. 80GSFC21M0002. We thank the anonymous referee for helpful comments and suggestions that improved the paper.

*Software:* cGENIE (https://github.com/derpycode/cgenie.muffin; Ridgwell et al. 2007), Atmos (https://github.com/VirtualPlanetaryLaboratory/atmos; Arney et al. 2016; Lincowski et al. 2018); Planetary Spectrum Generator (https://psg.gsfc.nasa.gov/; Villanueva et al. 2018); matplotlib (https://matplotlib.org/; Hunter 2007); numpy (https://numpy.org; Harris et al. 2020).

## Appendix A

Figure A1 displays the surface $N_2O$ mixing ratios in the same format as Figures 7 and 8, but for an Earthlike planet with $pO_2 = 10\%$ PAL.

Figure A2 shows the balance between the photolysis ($N_2O + h\nu$ [$\lambda < 240$ nm] $\rightarrow N_2 + O(^1D)$)) and radical ($N_2O + (O^1D) \rightarrow N_2 + O_2$ and $N_2O + O(^1D) \rightarrow NO + NO$) destruction channels as a function of the $N_2O$ mixing ratio for four potential stellar hosts (TRAPPIST-1 [M8V], AD Leonis [M3.5Ve], HD 85512 [K6V], and the Sun [G2V]), with $pO_2 = 100\%$ PAL. Loss rates are calculated for the corresponding scenarios in Figure 6.

Figure A3 shows a sensitivity test where a subset of the simulations from Figure 6 were run with different total surface pressures of 0.5, 1, and 10 bar. The $N_2$ partial pressure was adjusted to attain these different total pressures, while the total $O_2$ partial pressures were fixed at 100% PAL (i.e., 0.21 bars), to isolate the effects of pressure structure from the total $O_2$ column densities. The subset of stellar spectra chosen spans most of the main sequence and encompasses the minima and maxima of the observed flux–abundance relationships. To facilitate direct intercomparison, the $N_2O$ mixing ratios were adjusted to reflect the effective mixing ratio if the total pressure were 1 bar (e.g., the 10 bar surface mixing ratios were multiplied by 10). Simulations with a total pressure of 0.5 bar show very slight reductions in $N_2O$ at a given flux compared to the corresponding 1 bar case. Simulations with total pressures of 10 bar show enhanced $N_2O$ abundances at a given flux, by a larger factor of up to ~2–3, most notably for the Earth–Sun cases. The $N_2O$ abundance enhancements at high pressure were due to the adjustment of the altitudes where $O(^1D)$ was formed. In the 10 bar cases, $O(^1D)$, the secondary sink for $N_2O$, was most effectively formed in the upper atmosphere, where $N_2O$ concentrations were lower due to photolysis, reducing the impact of this photochemical sink. Correspondingly, the largest pressure effect is seen in the Sun cases, where $O(^1D)$ is a proportionally larger sink than it is for other stellar types (see Figure A2). This sensitivity test demonstrates that high-pressure atmospheres will facilitate enhanced $N_2O$ abundances at fixed $O_2$ partial pressures and fixed $N_2O$ fluxes. Importantly, the general slopes of the flux–abundance relationships for each star are robust to differences in the partial pressures of inert gases.

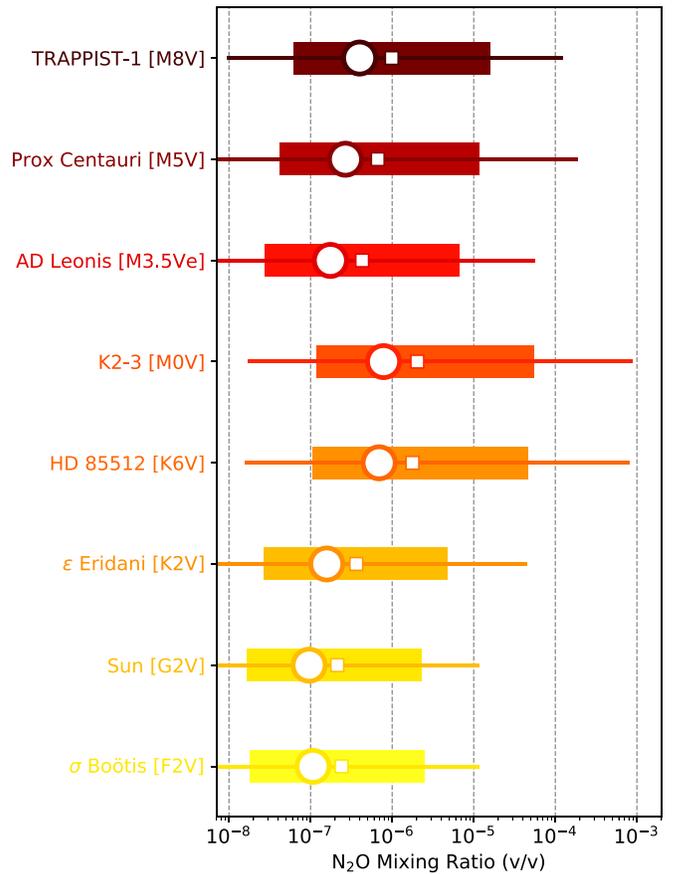

**Figure A1.** The same as Figures 7 and 8, but for an Earthlike planet with $pO_2 = 10\%$ PAL.

Figure A4 shows a sensitivity test where a subset of the simulations from Figure 6 were run with an eddy diffusion ($k_{zz}$) profile increased by a factor of 10 from the default Earth values above the troposphere. (This results in a shift from an average of $\sim 10^4$–$10^5$ cm$^{-2}$ s$^{-1}$ at 30 km). The cases with an increased eddy diffusion profile have lower concentrations by a maximum of a factor of ~3, due to more effective transport to the upper atmosphere, where $N_2O$ can be more easily photolyzed. The sensitivity of the $N_2O$ concentrations to the assumed $k_{zz}$ profile is substantially reduced for the later stellar types (K6V and M5V) versus the Sun, due to the difference in the spectral distribution of stellar UV photons between the stars and the increased importance of $N_2O$ self-shielding at higher $N_2O$ concentrations. The slopes of the flux–abundance relationships for each star are robust to uniform increases in assumed $k_{zz}$.

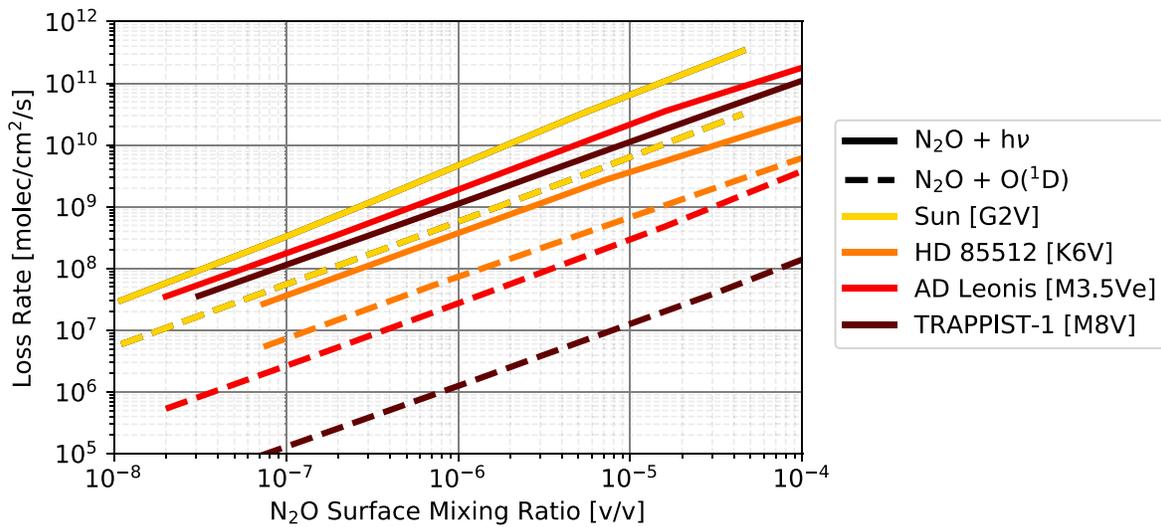

**Figure A2.** Photochemical reaction rates for the two major $N_2O$ destruction pathways as a function of stellar type. The solid lines represent the $N_2O$ photolysis rates, while the dotted lines represent the destruction of $N_2O$ by singlet oxygen $O(^1D)$. The calculations assume $pO_2 = 100\%$ PAL.

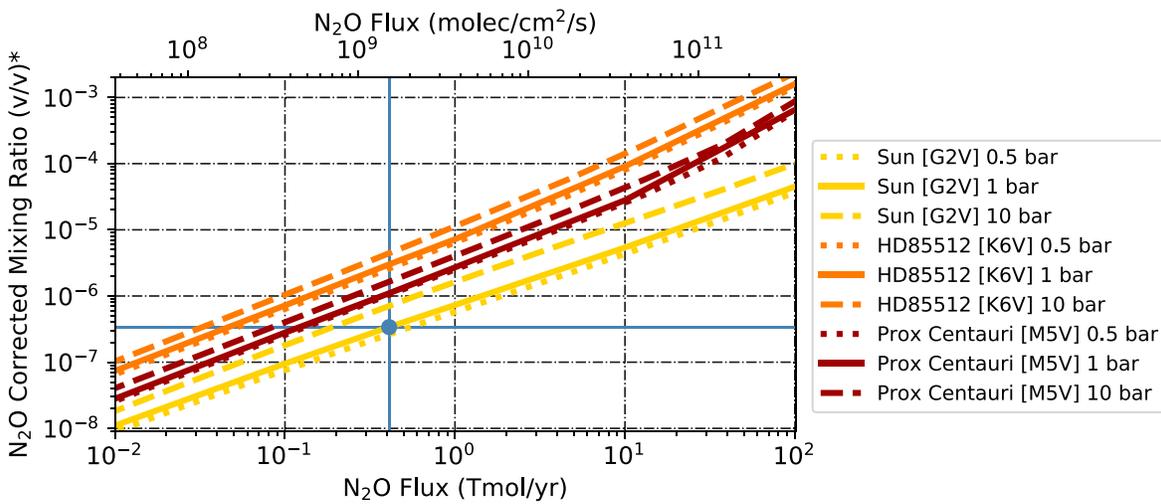

**Figure A3.** Sensitivity test showing the atmospheric mixing ratios of $N_2O$ as a function of surface flux for an Earthlike atmosphere with $pO_2 = 100\%$ PAL and total pressures of 0.5 bar (dotted lines), 1 bar (solid lines), and 10 bar (dashed lines). The total pressures are adjusted by changing the $N_2$ partial pressures, while the $O_2$ partial pressure is fixed at 0.21 bar. $N_2O$ mixing ratios are recalculated to reflect the effective mixing ratio if the total pressure were 1 bar.

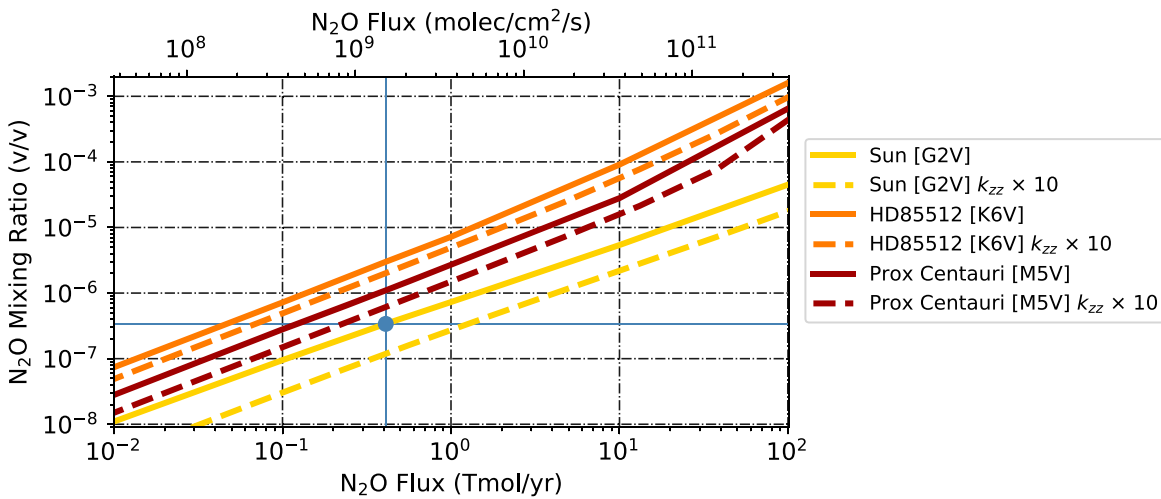

**Figure A4.** Sensitivity test showing the atmospheric mixing ratios of $N_2O$ as a function of surface flux for an Earthlike atmosphere with $pO_2 = 100\%$ PAL and an eddy diffusion ($k_{zz}$) profile matching the Earth average (solid) and increased by a factor of 10 uniformly through the upper atmosphere (dashed).

## ORCID iDs

Edward W. Schwieterman https://orcid.org/0000-0002-2949-2163
Stephanie L. Olson https://orcid.org/0000-0002-3249-6739
Daria Pidhorodetska https://orcid.org/0000-0001-9771-7953
Thomas J. Fauchez https://orcid.org/0000-0002-5967-9631
Sandra T. Bastelberger https://orcid.org/0000-0003-2052-3442
Jaime S. Crouse https://orcid.org/0000-0003-2273-8324
Andy Ridgwell https://orcid.org/0000-0003-2333-0128